\documentclass[%
 reprint,
 amsmath,amssymb,
 aps,
]{revtex4-1}

\usepackage[dvipsnames]{xcolor}

\usepackage{graphicx}
\usepackage{dcolumn}
\usepackage{bm}

\begin{document}

\preprint{APS/123-QED}

\title{Swirling against the forcing: evidence of stable counter-directed sloshing waves in orbital-shaken reservoirs}

\author{Alice Marcotte, François Gallaire}
 \email{francois.gallaire@epfl.ch}
 \author{Alessandro Bongarzone}
\affiliation{%
 Laboratory of Fluid Mechanics and Instabilities, École Polytechnique Fédérale de Lausanne, Lausanne, CH-1015, Switzerland
}%

\date{\today}

\begin{abstract}
We study the free surface response in a cylindrical container undergoing an elliptic periodic orbit. For small forcing amplitudes and deep liquid layers, we quantify the effect of orbit's aspect ratio onto the surface dynamics in the vicinity of the fluid system’s lowest natural frequency $\omega_0$. We provide experimental evidences of the existence of a frequency range where stable swirling can be either co- or counter-directed with respect to the container’s direction of motion. Our findings are successfully predicted by an inviscid asymptotic model, amended with a heuristic damping.
\end{abstract}

\maketitle


The problem of liquid sloshing pertains to many aspects of daily life, ranging from mundane wine tasting to more pragmatic issues such as liquid spilling \citep{mayer2012walking} and transport safety \citep{faltinsen2005liquid}. A proper predictive understanding and modelling of the sloshing hydrodynamics at stake is therefore essential in the design process of liquid tanks and in implementing control systems of vehicles \citep{ibrahim2009liquid}.\\
\indent In this regard, the case of orbital sloshing in partially filled circular cylinders represents one of the archetypal sloshing systems \citep{abramson1966dynamic}. Previous experimental studies have described its close-to-resonance dynamics for either circular or purely longitudinal shaking, casting light on a rich variety of wave regimes attracting interest to dynamicists over the last decades \citep{hutton1963inv,ockendon1973resonant,miles1984resonant,miles1984resonantly}.\\
\indent For circular orbits, the system responds with a swirling wave always co-directed with the container motion \citep{reclari2014surface}. This well-defined hydrodynamics, often simply modelled by a one-degree-of-freedom Duffing oscillator \citep{ockendon1973resonant,bongarzone2022amplitude}, is advantageously exploited, for example in biology, in the design of bioreactors, where the container is shaken so as to mix the liquid, prevent sedimentation and enhance gas transfer, hence providing a suitable oxygenation to the growing cell population \citep{klockner2012advances}. In the case of longitudinal forcing, the standing wave solution may undergo close-to-resonance a symmetry-breaking, with clock- and anti-clockwise swirling waves equally probable, or completely lose regularity showing an irregular and chaotic alternance of planar and swirling motions \citep{hutton1963inv,royon2007liquid}. Such a configuration finds a close mechanical analogy in the resonant motion of a forced spherical pendulum \citep{miles1984resonantly}, a four degrees-of-freedom system that has been widely studied in the context of order-to-chaos transitions \citep{miles1984resonant,tritton1986ordered} for its similarities with the Lorentz's problem \citep{lorenz1963deterministic}.\\
\indent Surprisingly however, no experimental studies devoted to the more generic case of elliptic orbits have been reported so far in the sloshing literature. Yet, existing theoretical analyses of this forcing condition brought out interesting features of the resonant liquid response that depend on the orbit’s ellipticity. In particular, a recent inviscid theory \cite{faltinsen2016resonant} suggested the counter-intuitive existence, under resonant elliptic forcing, of stable swirling waves that propagate in the direction \textit{opposite} to the forcing direction. Moreover, the theory anticipated that such counter-waves may exist even for quasi-circular orbits and travel with a smaller amplitude than co-directed waves. This, if confirmed, would further enrich the variety of observable dynamical sloshing regimes and possibly open new room in this rich dynamic system field.\\
\indent For moderately large-size containers, the use of inviscid hydrodynamic models is well accepted \citep{ibrahim2009liquid}, but in real sloshing problems, waves are always subjected to a non-vanishing viscous dissipation. Hence, the counter-swirling wave predicted by the inviscid model \cite{faltinsen2016resonant}, being intrinsically disfavored by the forcing direction, is likely to be more sensitive to damping than co-swirling solutions, and it is currently unclear whether such a solution can actually arise in a real-life lab-experiment.\\
\indent With this Letter, we aim at providing a joint experimental and theoretical characterization of the free liquid surface response for a generic, elliptic periodic container trajectory, so as to bridge the gap between the two diametrically opposed shaking conditions previously discussed. Specifically, we intend to identify the range of external control parameters, i.e. driving frequency, amplitude and orbit aspect ratio, for which stable counter-directed swirling waves do occur, and assess the extent of the forcing regime where inviscid models break down.\\
\indent To this end, we used a Plexiglas circular cylindrical container of total height $50\,\text{cm}$ and internal radius $R=0.086\,\text{m}$, filled to a depth $h=0.15\,\text{m}$ with water: density $\rho=1000\,\text{kg\,m$^{-3}$}$, surface tension $\gamma=0.0725\,\text{N\,m$^{-1}$}$ and dynamic viscosity $\mu=0.001\,\text{kg\,m$^{-1}$\,s$^{-1}$}$. 
\begin{figure}[t!]
\includegraphics[width=0.48\textwidth]{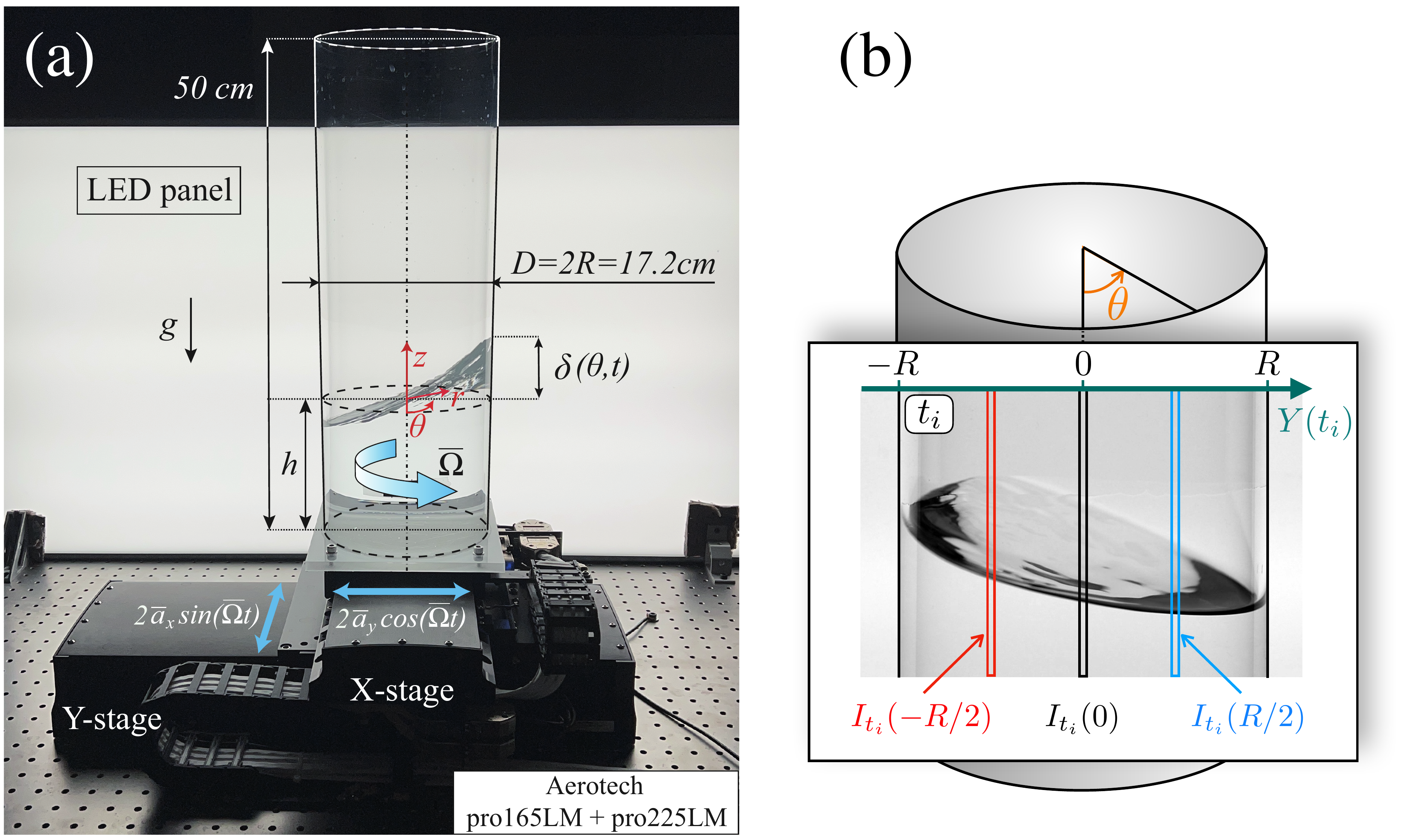}
\caption{(a) Experimental setup. Sloshing waves are generated by the container elliptic trajectory, achieved by imposing along the x and y axes two sinusoidal forcing of driving angular frequency $\overline{\Omega}$ and amplitudes $\overline{a}_x$ and $\overline{a}_y$. $\delta\left(\theta,t\right)$ denotes the free surface elevation measured at the sidewall, $r=R$. (b) Sketch illustrating the extraction from the frame corresponding to time-instant $t_i$ of the intensity profiles along the vertical middle axis of the container image (labelled as $I_{t_i}(0)$) and along the vertical axes located at coordinates $\mp R/2$.}
\label{fig:Fig1} 
\end{figure}
The gravity acceleration is denoted by $g$. The container is fixed on a double-axes linear motion actuator (Aerotech pro165LM + pro225LM), which imposes along the x and y axes two sinusoidal forcings of angular frequency $\overline{\Omega}$ and amplitudes $\overline{a}_x$ and $\overline{a}_y$, that are $\pi/2$-phase shifted with respect to each other. The fluid motion is recorded with a digital camera (Nikon D850) coupled with a Nikon 60mm f/2.8D lens and operated in slow mode with an acquisition frequency of 120fps. The camera optical axis is aligned with the x axis (see also Fig.~\ref{fig:Fig1}(a)). A LED panel is placed behind the container so as to provide a back illumination for a better optical contrast.\\
\indent In the moving reference frame, any planar elliptic-like shaking can be represented by the following equations describing the motion acceleration of the container axis parametrized in polar coordinates ($r$, $\theta$),
\begin{equation}
\label{eq:EqMotWall_ELL}
\frac{\text{d}^2\mathbf{X}_0}{\text{d}t^2}=
  \begin{cases}
    \left(-f_x\cos{\Omega t}\cos\theta-f_y\sin{\Omega t}\sin\theta\right)\,\mathbf{e}_r,\\
   \left(\, \, \, \, f_x\cos{\Omega t}\sin\theta-f_y\sin{\Omega t}\sin\theta\right)\,\mathbf{e}_{\theta},
  \end{cases}
\end{equation}
\noindent where $f_x=f=\overline{a}_x\Omega^2/R$ and $f_y=\alpha f=\overline{a}_y\Omega^2/R$ are the non-dimensional major- and minor-axis driving acceleration components, respectively, and $\Omega=\overline{\Omega}/\sqrt{g/R}$ the non-dimensional driving angular frequency. The bar symbol refers to the dimensional quantities. Note that the minor-to-major-axis component ratio, $\alpha=\overline{a}_y/\overline{a}_x=f_y/f_x$, has been introduced. A value $0<\alpha<1$ refers to elliptic orbits, whereas the two limiting cases $\alpha=0$ and $\alpha=1$ correspond, respectively, to longitudinal and circular shaking conditions.\\
\indent With this experimental campaign we intend to study the free surface response in the vicinity of the lowest natural frequency $\omega_0=\overline{\omega}_0/\sqrt{g/R}=\sqrt{k\tanh{\left(kh/R\right)}}=1.3547$ (with wavenumber $k=1.8412$) \citep{Lamb32}, for varying orbit's aspect ratios $\alpha$ and forcing amplitudes $\overline{a}_x$. In particular, we aim at triggering a counter-directed wave, if such a solution can arise in our experiments, and study how its existence depends on the forcing parameters.\\ 
\indent In a typical experiment, the amplitude $\overline{a}_x\in\left[1,3\right]\,\text{mm}$ and ellipticity $\alpha\in\left[0.10.95\right]$ are fixed, while frequencies are swept up- and downward within the (dimensionless) range $\Omega/\omega_0 \in [0.82, 1.21]$. The increment between two consecutive steps in the frequency sweep is $0.0217$. Each frequency step consists in two parts: the container undergoes first an harmonic elliptic forcing that is in the anti-clockwise direction for 150 oscillation periods and then in the clockwise direction for another 150 oscillation periods. Two movies are then recorded at each step so as to monitor the free surface response to both clockwise and anti-clockwise forcing. Switching the direction of the tank's trajectory induces a flow perturbation that can be enough to produce a counter-directed wave if the latter is an admissible configuration for the system. To ensure that the steady-state regime is established at each step and for each container direction, the camera is triggered only after 100 cycles so that it only records the last 50 oscillation periods.\\
\begin{figure}
\centering
\includegraphics[width=0.475\textwidth]{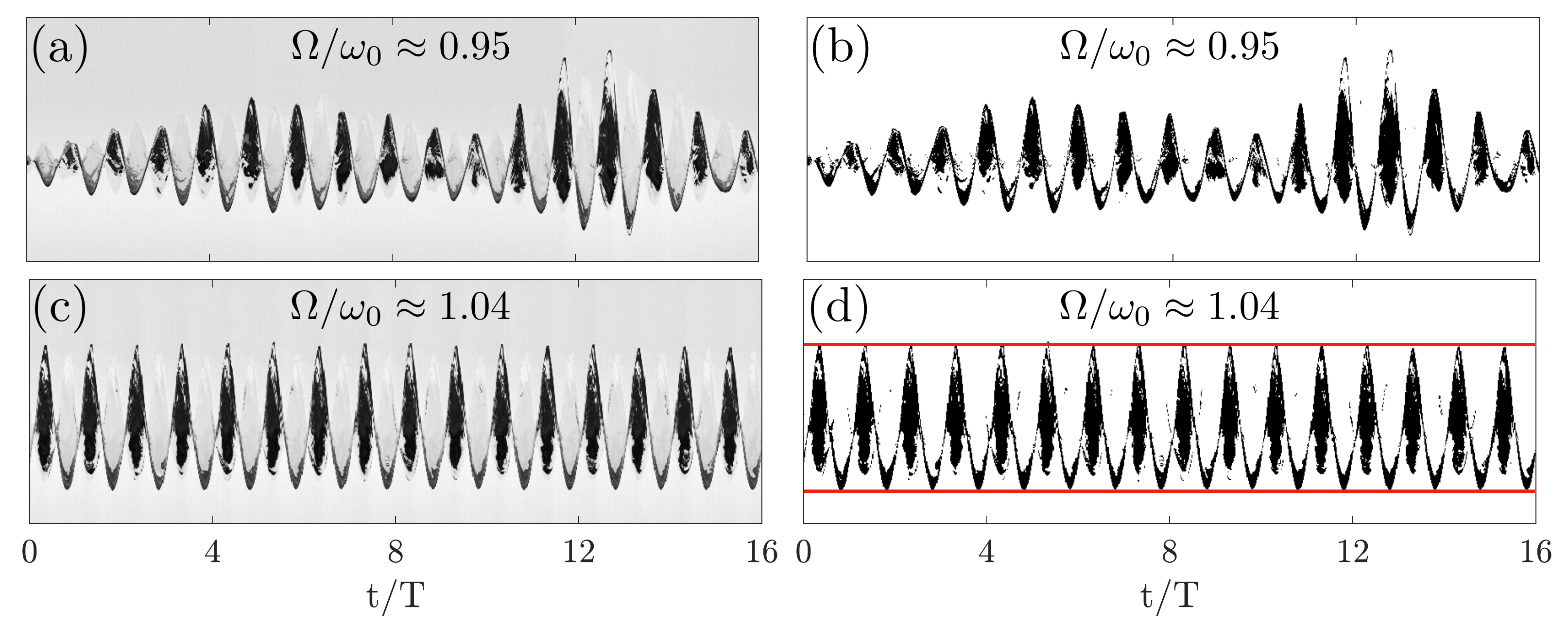}
\caption{(a)-(d) Intensity profiles as a function of time along the middle vertical axis $I(0)$ for ellipticity $\alpha=0.5$, amplitude $\overline{a}_x=1.5\,\text{mm}$ and frequency (a)-(b) $\Omega/\omega_0 \approx 0.95$ or (c)-(d) $\Omega/\omega_0 \approx 1.04$. The intensity profiles (b) and (d) are obtained from the binarization of (a) and (c) so as to filter out the signal of weaker intensity coming from the back contact line whenever the elevation of the front contact line is minimal. The oscillations of the front contact line are then enclosed into a top-bottom envelope, plotted in red in panel (d).}
\label{fig:regular_vs_irregular} 
\end{figure}
The procedure to analyze the free surface response is extensively described in \cite{marcotte2023super} and illustrated here in Fig.~\ref{fig:Fig1}(b). Briefly, we build from each movie an image $I(y)=[I_{t_1}(y), I_{t_2}(y), ...]$ where $I_{t_i}(y)$ is the intensity profile along the vertical axis $Y(t_i)=y$ on the frame $i$ corresponding to time $t_i$, with $Y(t_i)=0$ being the vertical middle axis between the edges of the container image (represented by $Y(t_i)=R$ and $Y(t_i)=-R$). The resulting image, as illustrated in Fig.~\ref{fig:regular_vs_irregular}, displays a periodic dark pattern that represents the free surface response to the harmonic forcing. The free surface appears as the darkest feature on each frame, so that the intensity profile along a vertical line at a given time $t_i$ represents the vertical extension of the free surface in this direction.\\ 
\begin{figure}[b!]
\centering
\includegraphics[width=0.485\textwidth]{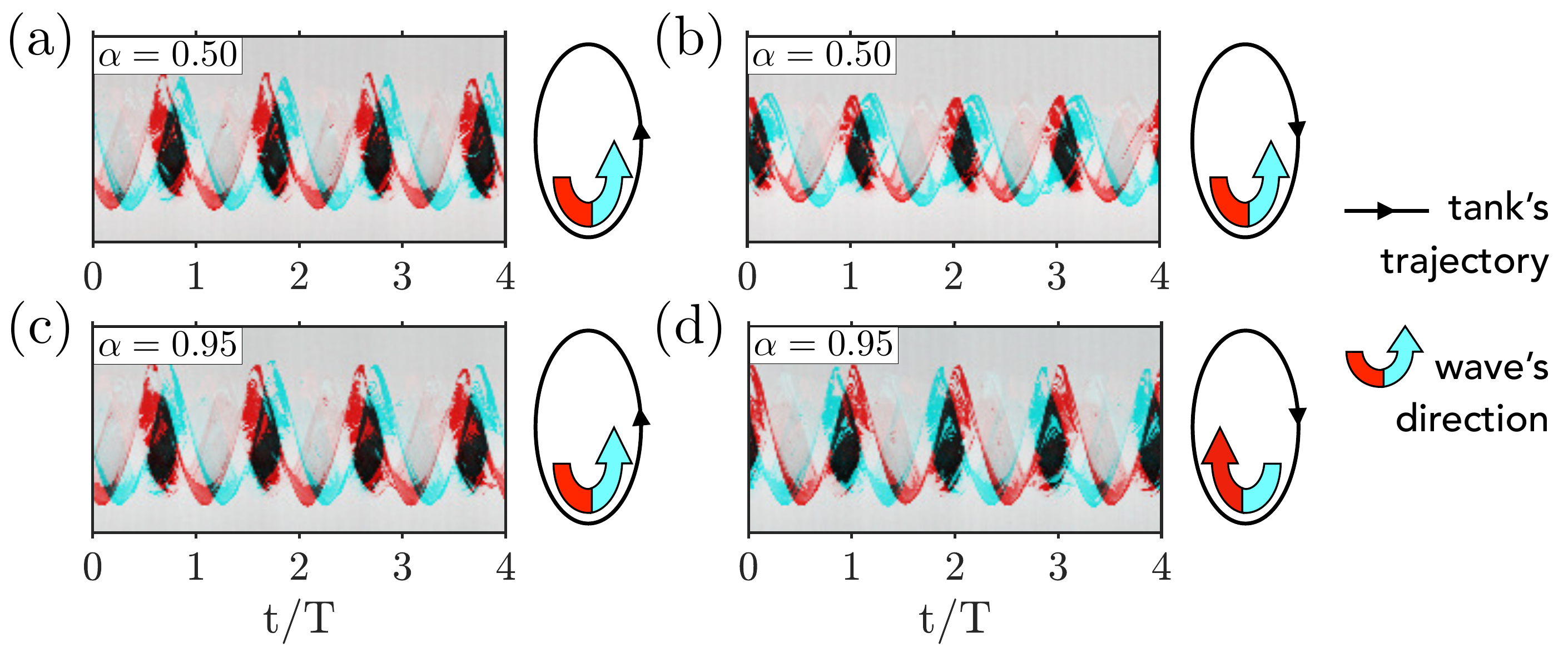}
\caption{Superposition of the intensity profiles as a function of time along the vertical axis $(Y=R/2)$ and $(Y=-R/2)$, denoted $I(R/2)$ (in blue) and $I(-R/2)$ (in red) respectively, for an harmonic forcing of frequency $\Omega/\omega_0 \approx 1.04$, amplitude $\overline{a}_x=1.5\,\text{mm}$, and (a)-(b) ellipticity $\alpha=0.50$ and (c)-(d) $\alpha = 0.95$. The container moves either in the anti-clockwise direction ((a) and (c)), or in the clockwise direction ((b) and (d)). See also Supplementary Movies \citep{SuppMov}.}
\label{fig:counter_rotating_waves} 
\end{figure}
\begin{figure}[t!]
\centering
\includegraphics[width=0.475\textwidth,height=0.3\textwidth]
{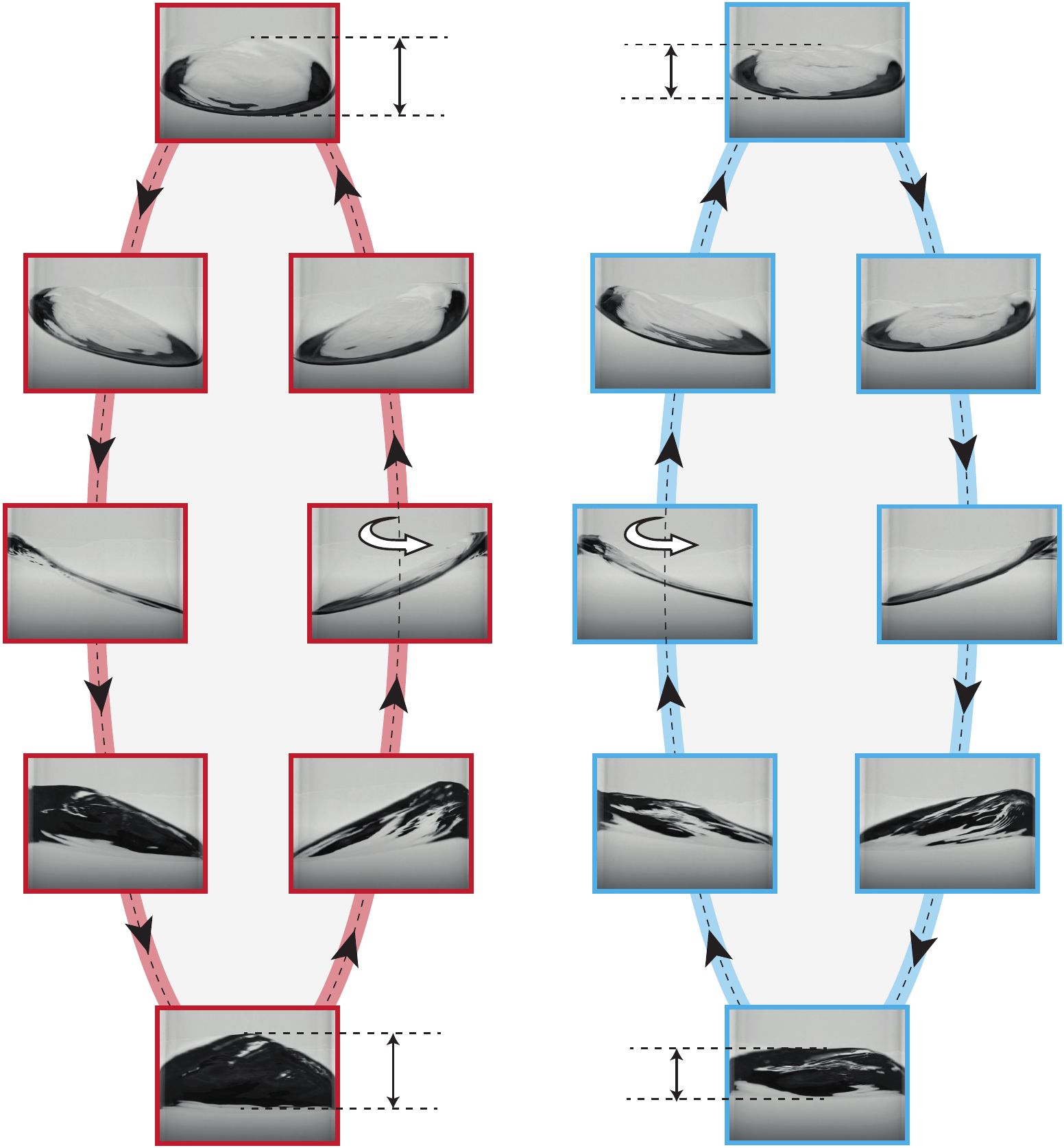}\\
\caption{Free surface snapshots corresponding to the case of Fig.~\ref{fig:counter_rotating_waves}(a)-(b) with $\alpha=0.50$. Direction of the container motion: \textit{left}, anti-clockwise; \textit{right}, clockwise (follow the black arrows). The white arrows indicate the direction of the wave rotation. A visual indication of the different wave amplitudes is provided by the black double-sided arrows.}
\label{fig:Fig5} 
\end{figure}
The usefulness of the resulting image $I(y)$ is threefold: (i) it allows to detect irregular dynamics. This corresponds to the absence of any stable wave-amplitude for a given set of forcing parameters, and is easily identified by the time-varying envelope modulating the free surface oscillations, see Fig.~\ref{fig:regular_vs_irregular}(a). (ii) For a regular response, $I(0)$ enables one to measure the amplitude of the front contact line in the azimuthal direction $\theta=0$. (iii) The comparison of the profiles along two vertical directions that are mirror-symmetric with respect to the vertical middle axis, e.g. $I(-R/2)$ and $I(R/2)$, makes possible to determine the propagation direction of the wave and to compare it with the container's motion direction.\\
\indent Fig.~\ref{fig:regular_vs_irregular} displays two intensity profiles as a function of time along the vertical middle axis $(Y=0)$ for same forcing amplitude $\overline{a}_x$ and ellipticity $\alpha$, but for two different forcing frequencies $\Omega/\omega_0$. Those images show that depending of the forcing parameters, the amplitude of the free surface oscillations can be either irregular, Fig.~\ref{fig:regular_vs_irregular}(a)-(b), or stationary, Fig.~\ref{fig:regular_vs_irregular}(c)-(d). In the analysis of the close-to-resonance dynamics, we therefore use the profile $I(0)$ to identify the irregular regime.\\ 
\begin{figure*}
\centering
\includegraphics[width=0.96\textwidth]{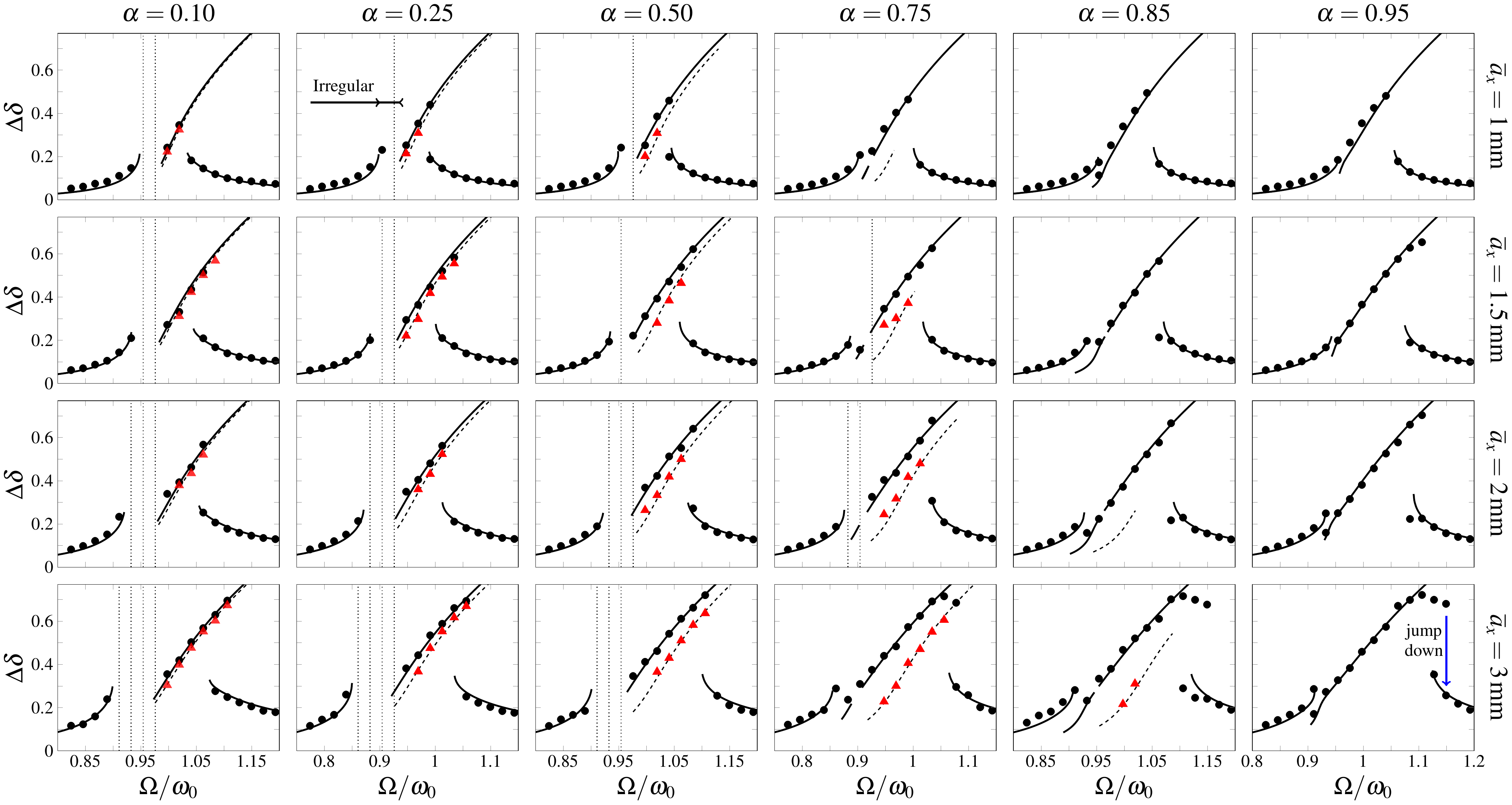}
\caption{Non-dimensional wave amplitude, $\Delta\delta=\left(\max_{t}\delta\left(0,t\right)-\min_{t}\delta\left(0,t\right)\right)/2R$ versus $\Omega/\omega_0$ for different values of $\overline{a}_x$ (columns) and $\alpha$ (rows). Markers: experiments (black for co- and and red for counter-waves). The typical dispersion in the measurements is well represented by the size of the markers. Curves: stable branches predicted by the present WNL theory (solid for co- and dashed for counter-waves). Vertical dotted lines indicate frequency values at which experiments have shown irregular motion. Unstable branches are not displayed for the sake of clarity.}
\label{fig:Fig2} 
\end{figure*}
The intensity profile $I(0)$ also provides the amplitude $\delta(\theta=0,t)$ of the swirling wave at the front wall of the container, i.e. at the azimuthal coordinate $\theta=0$, such as defined in Fig.~\ref{fig:Fig1}(b). Indeed, due to the back lighting, the intensity signal corresponding to the front contact line appears darker than the one due to the back contact line, so that informations associated with the latter can be filtered out by a proper thresholding of profile $I(0)$. On the resulting binarized image, the maximal and minimal heights of the final periodic pattern correspond then to the peaks and troughs of the swirling wave at the front wall along $\theta=0$. The amplitude of the wave (in pixel) is thus experimentally retrieved as half the difference between the height of the top and of the bottom envelopes enclosing its oscillations, displayed in in Fig.~\ref{fig:regular_vs_irregular}(d) as red lines, and converted into millimeters by using a scale put on the front wall of the container. Note that in this procedure, we neglect the variation of the pixel size that can occur along the container motion, the camera being fixed. This is justified by the very small forcing amplitude ($1\,\text{mm} \le\overline{a}_x\le 3\,\text{mm}$) with respect to the distance between the camera and the front wall of the container ($1\,\text{m}$). The error related to the variation of the pixel size is therefore of the order of $0.1\% $, i.e. negligible compared to the typical dispersion of our measurements.\\
\indent To detect the direction of propagation of the wave, we compare for each movie the intensity profiles along two vertical directions that are mirror-symmetric with respect to the vertical middle axis of the container. Fig.~\ref{fig:counter_rotating_waves} shows composite images that each consists in the superposition of $I(R/2)$ and $I(-R/2)$ into a composite RGB image, where gray areas correspond to pixels where $I(R/2)$ and $I(-R/2)$ have the same intensity, while red (resp. blue) areas correspond to the part of $I(-R/2)$ (resp. $I(R/2)$) that do not overlap with $I(R/2)$ (resp. $I(-R/2)$). Thus, a red (resp. blue) peak preceding a blue (resp. red) peak corresponds to a wave traveling from the left (resp. right) to the right (resp. left) hand-side of the front wall of the container, i.e. in the anti-clockwise (resp. clockwise) direction. The propagation direction of the wave can then be determined and compared to the direction of the container motion. In Fig.~\ref{fig:counter_rotating_waves}, the dynamics associated with two different aspect ratio $\alpha=0.5$ and $\alpha = 0.95$ (quasi-circular orbit) are compared for the same forcing frequency and amplitude. For each $\alpha$, the right and left hand-side signals $I(R/2)$ and $I(-R/2)$ are superposed to each other for two motion configurations, namely an anti-clockwise followed by a clockwise container trajectory. In the case of the anti-clockwise tank's motion, Fig.~\ref{fig:counter_rotating_waves}(a)-(b), the swirling wave travels in the same direction as the container, but the change of container's motion direction induces a flow perturbation sufficient to produce a robust counter-directed wave, if the latter corresponds to a system's stable solution. We indeed observe in the case of $\alpha = 0.5$ that the wave, though of smaller amplitude, is still traveling from the left to the right hand-side of the container's front wall despite the reverse of direction in the tank trajectory. This appears glaring in Fig.~\ref{fig:Fig5}, where the two series of free surface snapshots show how the wave's direction of rotation remains unchanged despite the reversal of the container's direction of motion. On the contrary, Fig.~\ref{fig:counter_rotating_waves}(c)-(d), the wave switches direction for the large ellipticity $\alpha =0.95$ and is therefore co-directed with the forcing for both container motion directions. These results provide the first experimental evidence for the existence of counter-swirling solutions, and validate this procedure as suitable to trigger and identify stable counter-directed waves.\\
\indent To assess the extent of the validity of an inviscid, and thus simplified, hydrodynamic model to predict resonant counter-swirling in a lab-scale experiment, we aim at comparing our experimental data to quantitative predictions provided by the model formalized in \cite{marcotte2023super} and recalled in the Supplementary Materials \citep{SuppMat}. This weakly nonlinear model has been extensively compared with \cite{faltinsen2016resonant} for both purely circular \cite{bongarzone2022sub} and longitudinal \cite{marcotte2023super} shaking conditions and it has been shown to provide consistent results. Briefly, we assume an inviscid, irrotational and incompressible flow (as common in sloshing theory applied to lab-scale containers \citep{faltinsen2005liquid,ibrahim2009liquid}) and a small forcing acceleration $f \sim \epsilon^3$, with $\epsilon$ a small parameter $\ll 1$. This last assumption is justified by the fact that close to the resonance $\Omega \approx \omega_0$, even a small forcing will induce a large system response. Our model consists then in a multiple time-scale weakly nonlinear (WNL) analysis \cite{whitham1974linear}, whose lowest order asymptotic solution is described by the following system of amplitude equations governing the close-to-resonance interaction of two counter-propagating waves of complex amplitudes $A=|A|e^{\text{i}\Phi_A}$ $\left(-\right)$ and $B=|B|e^{\text{i}\Phi_B}$ $\left(+\right)$ and oscillating in time-space as $\cos{\left(\Omega t \mp\theta\right)}$, such that the free surface elevation $\delta\left(0,t\right)$, e.g. measured at the lateral wall, $r=R$, and at an azimuthal coordinate  $\theta=0$, reads $\delta\left(0,t\right)/R=2\left(|A|\cos{\left(\Omega t +\Phi_A\right)}+|B|\cos{\left(\Omega t+\Phi_B\right)}\right)$, 
\begin{subequations}
\begin{equation}
\label{eq:AmpEqSCfinalA_elliptic}
\frac{\text{d}A}{\text{d}t}=-\left(\sigma+\text{i}\lambda\right) A + \text{i}\,\mu \alpha_{_A} f + \text{i}\,\nu |A|^2A +\text{i}\,\xi |B|^2A,
\end{equation}
\begin{equation}
\label{eq:AmpEqSCfinalB_elliptic}
\frac{\text{d}B}{\text{d}t}=-\left(\sigma+\text{i}\lambda\right) B + \text{i}\,\mu \alpha_{_B} f + \text{i}\,\nu |B|^2B + \text{i}\,\xi |A|^2B.
\end{equation}
\end{subequations}
\noindent The parameter $\lambda=\Omega-\omega_0$ represents a frequency detuning with respect to the exact resonance, whereas $\alpha_{_A}=\left(1+\alpha\right)/2$ and $\alpha_{_B}=\left(1-\alpha\right)/2$ are two auxiliary orbit parameters. The values of $\mu$, $\nu$ and $\xi\in\mathbb{R}$ are given in Supplementary Material \citep{SuppMat}.\\
\indent In contradistinction with \cite{marcotte2023super}, here we account for an heuristic viscous damping coefficient, $\sigma$, introduced \textit{a posteriori} in~\eqref{eq:AmpEqSCfinalA_elliptic}-\eqref{eq:AmpEqSCfinalB_elliptic} and whose value is estimated according to equation~(17) of \cite{raynovskyy2018damped}. For our setup, the damping associated with $\omega_0$ amounts to $\sigma=0.0055$.\\ 
\indent We now compare, in Fig.~\ref{fig:Fig2}, our measurements to the asymptotic model~\eqref{eq:AmpEqSCfinalA_elliptic}-\eqref{eq:AmpEqSCfinalB_elliptic}. At small ellipticity, e.g. $\alpha$ close to $0.10$, the amplitude response curve is similar to that induced by a longitudinal forcing \citep{royon2007liquid,marcotte2023super} except that the planar wave solution no longer exists, owing to the preferential direction of motion, and that the co- and counter-rotating waves are no more equally probable, with the counter-wave exhibiting a slightly lower amplitude. By increasing the value of $\alpha$, the counter-wave displays a decreasing amplitude and the range of frequency for which irregular motion occurs shrinks down and ultimately vanishes \citep{faltinsen2016resonant}. In this context, irregular means that both the co- and counter-swirling solutions are unstable and the system exhibits irregular and chaotic patterns switching between co- and, at a small ellipticity, counter-swirling dynamics alternating transient intervals of nearly-planar motions (see Supplementary Movies \citep{SuppMov}). As $\alpha$ approaches $1$, the admissible frequency range associated for counter-waves reduces and it is eventually suppressed, whereas the frequency range associated with co-directed swirling widens and covers all of the resonance frequency range at $\alpha=0.95$, i.e. approaching the limiting case of a circular trajectory ($\alpha=1$) \citep{reclari2014surface,bongarzone2022amplitude}. 
We also observe a decrease in the wave amplitude at $\overline{a}_x=3\,\text{mm}$ for $\alpha\ge 0.5$, occurring just before the jump-down frequency and which can be tentatively attributed to increasing nonlinear effects overlooked by the WNL model.\\
\indent The experimental steady-state wave amplitudes are in a good quantitative agreement with the WNL analysis for all $\overline{a}_x$ and $\alpha$ values explored, hence proving the validity of the inviscid analysis in our regime of operation. The only major limitation is intrinsic to the use of a simple phenomenological damping. As the latter does not depend on the wave amplitude, it cannot accurately predict the phase-lag between forcing and the system response \citep{bauerlein2021phase}. This translates in an imprecise estimation of the jump-down frequency occurring above resonance and of the frequency range associated to the counter-swirling, that appears slightly overestimated.\\
\indent To conclude, we have investigated the sloshing dynamics in the vicinity of the first harmonic resonance, i.e. $\Omega/\omega_0\approx1$, for container's elliptic orbits. The amplitude-response curves at different forcing amplitudes were examined versus the orbit aspect ratio, $0<\alpha<1$. Our experiments demonstrated the existence of a significant frequency range associated with stable counter-swirling, up to surprisingly high orbit's aspect ratio.\\
\indent Our findings have been successfully rationalized by the asymptotic model formalized in \cite{marcotte2023super} starting from the full hydrodynamic system supplemented with a simple heuristic damping coefficient. The close-to-resonant sloshing dynamics for any container's elliptic-like orbit is therefore well represented by four degrees of freedom only. This suggests that a simple generalization of the forcing condition in the problem of the resonantly forced spherical pendulum \citep{miles1984resonant} could provide a suitable mechanical analogy for this entire class of sloshing dynamics, thus offering further room in this archetypical low degrees-of-freedom class of dynamical systems.\\
\indent Perspectives include the experimental characterization of the effect of variations in the fluid depth and the WNL model extension to a viscous framework \citep{bongarzone2022sub}. Although the latter presently hinges on the subtle modelling of the moving contact line dynamics, such an extension is desirable, as it would enable one to better quantify the overall system dissipation and to predict the viscous streaming experimentally observed in orbital shaken containers \citep{bouvard2017mean}.\\

\indent This work was founded by the Swiss National Science Foundation under grants 178971 and 200341.

\bibliography{Bibliography}

\begin{thebibliography}{24}%
\makeatletter
\providecommand \@ifxundefined [1]{%
 \@ifx{#1\undefined}
}%
\providecommand \@ifnum [1]{%
 \ifnum #1\expandafter \@firstoftwo
 \else \expandafter \@secondoftwo
 \fi
}%
\providecommand \@ifx [1]{%
 \ifx #1\expandafter \@firstoftwo
 \else \expandafter \@secondoftwo
 \fi
}%
\providecommand \natexlab [1]{#1}%
\providecommand \enquote  [1]{``#1''}%
\providecommand \bibnamefont  [1]{#1}%
\providecommand \bibfnamefont [1]{#1}%
\providecommand \citenamefont [1]{#1}%
\providecommand \href@noop [0]{\@secondoftwo}%
\providecommand \href [0]{\begingroup \@sanitize@url \@href}%
\providecommand \@href[1]{\@@startlink{#1}\@@href}%
\providecommand \@@href[1]{\endgroup#1\@@endlink}%
\providecommand \@sanitize@url [0]{\catcode `\\12\catcode `\$12\catcode
  `\&12\catcode `\#12\catcode `\^12\catcode `\_12\catcode `\%12\relax}%
\providecommand \@@startlink[1]{}%
\providecommand \@@endlink[0]{}%
\providecommand \url  [0]{\begingroup\@sanitize@url \@url }%
\providecommand \@url [1]{\endgroup\@href {#1}{\urlprefix }}%
\providecommand \urlprefix  [0]{URL }%
\providecommand \Eprint [0]{\href }%
\providecommand \doibase [0]{https://doi.org/}%
\providecommand \selectlanguage [0]{\@gobble}%
\providecommand \bibinfo  [0]{\@secondoftwo}%
\providecommand \bibfield  [0]{\@secondoftwo}%
\providecommand \translation [1]{[#1]}%
\providecommand \BibitemOpen [0]{}%
\providecommand \bibitemStop [0]{}%
\providecommand \bibitemNoStop [0]{.\EOS\space}%
\providecommand \EOS [0]{\spacefactor3000\relax}%
\providecommand \BibitemShut  [1]{\csname bibitem#1\endcsname}%
\let\auto@bib@innerbib\@empty
\bibitem [{\citenamefont {Mayer}\ and\ \citenamefont
  {Krechetnikov}(2012)}]{mayer2012walking}%
  \BibitemOpen
  \bibfield  {author} {\bibinfo {author} {\bibfnamefont {H.~C.}\ \bibnamefont
  {Mayer}}\ and\ \bibinfo {author} {\bibfnamefont {R.}~\bibnamefont
  {Krechetnikov}},\ }\href@noop {} {\bibfield  {journal} {\bibinfo  {journal}
  {Phys. Rev. E}\ }\textbf {\bibinfo {volume} {85}},\ \bibinfo {pages} {046117}
  (\bibinfo {year} {2012})}\BibitemShut {NoStop}%
\bibitem [{\citenamefont {Faltinsen}\ and\ \citenamefont
  {Timokha}(2009)}]{faltinsen2005liquid}%
  \BibitemOpen
  \bibfield  {author} {\bibinfo {author} {\bibfnamefont {O.~M.}\ \bibnamefont
  {Faltinsen}}\ and\ \bibinfo {author} {\bibfnamefont {A.~N.}\ \bibnamefont
  {Timokha}},\ }\href@noop {} {\emph {\bibinfo {title} {Sloshing}}}\ (\bibinfo
  {publisher} {Cambridge University Press},\ \bibinfo {year}
  {2009})\BibitemShut {NoStop}%
\bibitem [{\citenamefont {Ibrahim}(2005)}]{ibrahim2009liquid}%
  \BibitemOpen
  \bibfield  {author} {\bibinfo {author} {\bibfnamefont {R.~A.}\ \bibnamefont
  {Ibrahim}},\ }\href@noop {} {\emph {\bibinfo {title} {Liquid sloshing
  dynamics: Theory and Applications}}}\ (\bibinfo  {publisher} {Cambridge
  University Press},\ \bibinfo {year} {2005})\BibitemShut {NoStop}%
\bibitem [{\citenamefont {Abramson}(1966)}]{abramson1966dynamic}%
  \BibitemOpen
  \bibfield  {author} {\bibinfo {author} {\bibfnamefont {H.~N.}\ \bibnamefont
  {Abramson}},\ }\href@noop {} {}\ (\bibinfo  {publisher} {\textit{NASA Tech.
  Rep.} SP-106. NASA, Washington},\ \bibinfo {year} {1966})\BibitemShut
  {NoStop}%
\bibitem [{\citenamefont {Hutton}(1963)}]{hutton1963inv}%
  \BibitemOpen
  \bibfield  {author} {\bibinfo {author} {\bibfnamefont {R.~E.}\ \bibnamefont
  {Hutton}},\ }\href@noop {} {}\ (\bibinfo  {publisher} {\textit{NASA Tech.
  Rep.} NASA; D-1870},\ \bibinfo {year} {1963})\BibitemShut {NoStop}%
\bibitem [{\citenamefont {Ockendon}\ and\ \citenamefont
  {Ockendon}(1973)}]{ockendon1973resonant}%
  \BibitemOpen
  \bibfield  {author} {\bibinfo {author} {\bibfnamefont {J.~R.}\ \bibnamefont
  {Ockendon}}\ and\ \bibinfo {author} {\bibfnamefont {H.}~\bibnamefont
  {Ockendon}},\ }\bibfield  {title} {\bibinfo {title} {Resonant surface
  waves},\ }\href@noop {} {\bibfield  {journal} {\bibinfo  {journal} {J.~Fluid
  Mech.}\ }\textbf {\bibinfo {volume} {59}},\ \bibinfo {pages} {397} (\bibinfo
  {year} {1973})}\BibitemShut {NoStop}%
\bibitem [{\citenamefont {Miles}(1984{\natexlab{a}})}]{miles1984resonant}%
  \BibitemOpen
  \bibfield  {author} {\bibinfo {author} {\bibfnamefont {J.~W.}\ \bibnamefont
  {Miles}},\ }\href@noop {} {\bibfield  {journal} {\bibinfo  {journal} {Phys.
  D: Nonlinear Phenom.}\ }\textbf {\bibinfo {volume} {11}},\ \bibinfo {pages}
  {309} (\bibinfo {year} {1984}{\natexlab{a}})}\BibitemShut {NoStop}%
\bibitem [{\citenamefont {Miles}(1984{\natexlab{b}})}]{miles1984resonantly}%
  \BibitemOpen
  \bibfield  {author} {\bibinfo {author} {\bibfnamefont {J.~W.}\ \bibnamefont
  {Miles}},\ }\href@noop {} {\bibfield  {journal} {\bibinfo  {journal}
  {J.~Fluid Mech.}\ }\textbf {\bibinfo {volume} {149}},\ \bibinfo {pages} {15}
  (\bibinfo {year} {1984}{\natexlab{b}})}\BibitemShut {NoStop}%
\bibitem [{\citenamefont {Reclari}\ \emph {et~al.}(2014)\citenamefont
  {Reclari}, \citenamefont {Dreyer}, \citenamefont {Tissot}, \citenamefont
  {Obreschkow}, \citenamefont {Wurm},\ and\ \citenamefont
  {Farhat}}]{reclari2014surface}%
  \BibitemOpen
  \bibfield  {author} {\bibinfo {author} {\bibfnamefont {M.}~\bibnamefont
  {Reclari}}, \bibinfo {author} {\bibfnamefont {M.}~\bibnamefont {Dreyer}},
  \bibinfo {author} {\bibfnamefont {S.}~\bibnamefont {Tissot}}, \bibinfo
  {author} {\bibfnamefont {D.}~\bibnamefont {Obreschkow}}, \bibinfo {author}
  {\bibfnamefont {F.~M.}\ \bibnamefont {Wurm}},\ and\ \bibinfo {author}
  {\bibfnamefont {M.}~\bibnamefont {Farhat}},\ }\href@noop {} {\bibfield
  {journal} {\bibinfo  {journal} {Phys. Fluids}\ }\textbf {\bibinfo {volume}
  {26}},\ \bibinfo {pages} {052104} (\bibinfo {year} {2014})}\BibitemShut
  {NoStop}%
\bibitem [{\citenamefont {Bongarzone}\ \emph
  {et~al.}(2022{\natexlab{a}})\citenamefont {Bongarzone}, \citenamefont
  {Guido},\ and\ \citenamefont {Gallaire}}]{bongarzone2022amplitude}%
  \BibitemOpen
  \bibfield  {author} {\bibinfo {author} {\bibfnamefont {A.}~\bibnamefont
  {Bongarzone}}, \bibinfo {author} {\bibfnamefont {M.}~\bibnamefont {Guido}},\
  and\ \bibinfo {author} {\bibfnamefont {F.}~\bibnamefont {Gallaire}},\
  }\href@noop {} {\bibfield  {journal} {\bibinfo  {journal} {J.~Fluid Mech.}\
  }\textbf {\bibinfo {volume} {943}},\ \bibinfo {pages} {A28} (\bibinfo {year}
  {2022}{\natexlab{a}})}\BibitemShut {NoStop}%
\bibitem [{\citenamefont {Kl{\"o}ckner}\ and\ \citenamefont
  {B{\"u}chs}(2012)}]{klockner2012advances}%
  \BibitemOpen
  \bibfield  {author} {\bibinfo {author} {\bibfnamefont {W.}~\bibnamefont
  {Kl{\"o}ckner}}\ and\ \bibinfo {author} {\bibfnamefont {J.}~\bibnamefont
  {B{\"u}chs}},\ }\href@noop {} {\bibfield  {journal} {\bibinfo  {journal}
  {Trends Biotechnol.}\ }\textbf {\bibinfo {volume} {30}},\ \bibinfo {pages}
  {307} (\bibinfo {year} {2012})}\BibitemShut {NoStop}%
\bibitem [{\citenamefont {Royon-Lebeaud}\ \emph {et~al.}(2007)\citenamefont
  {Royon-Lebeaud}, \citenamefont {Hopfinger},\ and\ \citenamefont
  {Cartellier}}]{royon2007liquid}%
  \BibitemOpen
  \bibfield  {author} {\bibinfo {author} {\bibfnamefont {A.}~\bibnamefont
  {Royon-Lebeaud}}, \bibinfo {author} {\bibfnamefont {E.~J.}\ \bibnamefont
  {Hopfinger}},\ and\ \bibinfo {author} {\bibfnamefont {A.}~\bibnamefont
  {Cartellier}},\ }\href@noop {} {\bibfield  {journal} {\bibinfo  {journal}
  {J.~Fluid Mech.}\ }\textbf {\bibinfo {volume} {577}},\ \bibinfo {pages} {467}
  (\bibinfo {year} {2007})}\BibitemShut {NoStop}%
\bibitem [{\citenamefont {Tritton}(1986)}]{tritton1986ordered}%
  \BibitemOpen
  \bibfield  {author} {\bibinfo {author} {\bibfnamefont {D.~J.}\ \bibnamefont
  {Tritton}},\ }\href@noop {} {\bibfield  {journal} {\bibinfo  {journal} {Eur.
  J. Phys.}\ }\textbf {\bibinfo {volume} {7}},\ \bibinfo {pages} {162}
  (\bibinfo {year} {1986})}\BibitemShut {NoStop}%
\bibitem [{\citenamefont {Lorenz}(1963)}]{lorenz1963deterministic}%
  \BibitemOpen
  \bibfield  {author} {\bibinfo {author} {\bibfnamefont {E.~N.}\ \bibnamefont
  {Lorenz}},\ }\href@noop {} {\bibfield  {journal} {\bibinfo  {journal} {J.
  Atmos. Sci.}\ }\textbf {\bibinfo {volume} {20}},\ \bibinfo {pages} {130}
  (\bibinfo {year} {1963})}\BibitemShut {NoStop}%
\bibitem [{\citenamefont {Faltinsen}\ \emph {et~al.}(2016)\citenamefont
  {Faltinsen}, \citenamefont {Lukovsky},\ and\ \citenamefont
  {Timokha}}]{faltinsen2016resonant}%
  \BibitemOpen
  \bibfield  {author} {\bibinfo {author} {\bibfnamefont {O.~M.}\ \bibnamefont
  {Faltinsen}}, \bibinfo {author} {\bibfnamefont {I.~A.}\ \bibnamefont
  {Lukovsky}},\ and\ \bibinfo {author} {\bibfnamefont {A.~N.}\ \bibnamefont
  {Timokha}},\ }\href@noop {} {\bibfield  {journal} {\bibinfo  {journal}
  {J.~Fluid Mech.}\ }\textbf {\bibinfo {volume} {804}},\ \bibinfo {pages} {608}
  (\bibinfo {year} {2016})}\BibitemShut {NoStop}%
\bibitem [{\citenamefont {Lamb}(1993)}]{Lamb32}%
  \BibitemOpen
  \bibfield  {author} {\bibinfo {author} {\bibfnamefont {H.}~\bibnamefont
  {Lamb}},\ }\href@noop {} {\emph {\bibinfo {title} {Hydrodynamics}}}\
  (\bibinfo  {publisher} {Cambridge University Press},\ \bibinfo {year}
  {1993})\BibitemShut {NoStop}%
\bibitem [{\citenamefont {Marcotte}\ \emph {et~al.}(2023)\citenamefont
  {Marcotte}, \citenamefont {Gallaire},\ and\ \citenamefont
  {Bongarzone}}]{marcotte2023super}%
  \BibitemOpen
  \bibfield  {author} {\bibinfo {author} {\bibfnamefont {A.}~\bibnamefont
  {Marcotte}}, \bibinfo {author} {\bibfnamefont {F.}~\bibnamefont {Gallaire}},\
  and\ \bibinfo {author} {\bibfnamefont {A.}~\bibnamefont {Bongarzone}},\
  }\href@noop {} {\bibfield  {journal} {\bibinfo  {journal} {arXiv preprint
  arXiv:2302.02443}\ } (\bibinfo {year} {2023})}\BibitemShut {NoStop}%
\bibitem [{Sup({\natexlab{a}})}]{SuppMov}%
  \BibitemOpen
  \href@noop {} {}\bibinfo {number} {{Supplementary Movies at [URL will be
  inserted by publisher] show an example of irregular motion as well as co- and
  counter-swirling dynamics experimentally observed for
  $\bar{a}_x=1.5\,\text{mm}$ at $\alpha=0.25$, $0.50$ and $0.95$}.}\BibitemShut
  {Stop}%
\bibitem [{Sup({\natexlab{b}})}]{SuppMat}%
  \BibitemOpen
  \href@noop {} {}\bibinfo {number} {{See Supplementary Material at [URL will
  be inserted by publisher] for details about the derivation of the asymptotic
  model, damping estimation and computation of stable stationary amplitudes
  from (3a)-(3b)}.}\BibitemShut {Stop}%
\bibitem [{\citenamefont {Bongarzone}\ \emph
  {et~al.}(2022{\natexlab{b}})\citenamefont {Bongarzone}, \citenamefont
  {Viola}, \citenamefont {Camarri},\ and\ \citenamefont
  {Gallaire}}]{bongarzone2022sub}%
  \BibitemOpen
  \bibfield  {author} {\bibinfo {author} {\bibfnamefont {A.}~\bibnamefont
  {Bongarzone}}, \bibinfo {author} {\bibfnamefont {F.}~\bibnamefont {Viola}},
  \bibinfo {author} {\bibfnamefont {S.}~\bibnamefont {Camarri}},\ and\ \bibinfo
  {author} {\bibfnamefont {F.}~\bibnamefont {Gallaire}},\ }\href@noop {}
  {\bibfield  {journal} {\bibinfo  {journal} {J.~Fluid Mech.}\ }\textbf
  {\bibinfo {volume} {947}},\ \bibinfo {pages} {A24} (\bibinfo {year}
  {2022}{\natexlab{b}})}\BibitemShut {NoStop}%
\bibitem [{\citenamefont {Whitham}(1974)}]{whitham1974linear}%
  \BibitemOpen
  \bibfield  {author} {\bibinfo {author} {\bibfnamefont {G.~B.}\ \bibnamefont
  {Whitham}},\ }\href@noop {} {\emph {\bibinfo {title} {Linear and nonlinear
  waves}}}\ (\bibinfo  {publisher} {Wiley, New York},\ \bibinfo {year}
  {1974})\BibitemShut {NoStop}%
\bibitem [{\citenamefont {Raynovskyy}\ and\ \citenamefont
  {Timokha}(2018)}]{raynovskyy2018damped}%
  \BibitemOpen
  \bibfield  {author} {\bibinfo {author} {\bibfnamefont {I.~A.}\ \bibnamefont
  {Raynovskyy}}\ and\ \bibinfo {author} {\bibfnamefont {A.~N.}\ \bibnamefont
  {Timokha}},\ }\href@noop {} {\bibfield  {journal} {\bibinfo  {journal} {Fluid
  Dyn. Res.}\ }\textbf {\bibinfo {volume} {50}},\ \bibinfo {pages} {045502}
  (\bibinfo {year} {2018})}\BibitemShut {NoStop}%
\bibitem [{\citenamefont {B{\"a}uerlein}\ and\ \citenamefont
  {Avila}(2021)}]{bauerlein2021phase}%
  \BibitemOpen
  \bibfield  {author} {\bibinfo {author} {\bibfnamefont {B.}~\bibnamefont
  {B{\"a}uerlein}}\ and\ \bibinfo {author} {\bibfnamefont {K.}~\bibnamefont
  {Avila}},\ }\href@noop {} {\bibfield  {journal} {\bibinfo  {journal}
  {J.~Fluid Mech.}\ }\textbf {\bibinfo {volume} {925}} (\bibinfo {year}
  {2021})}\BibitemShut {NoStop}%
\bibitem [{\citenamefont {Bouvard}\ \emph {et~al.}(2017)\citenamefont
  {Bouvard}, \citenamefont {Herreman},\ and\ \citenamefont
  {Moisy}}]{bouvard2017mean}%
  \BibitemOpen
  \bibfield  {author} {\bibinfo {author} {\bibfnamefont {J.}~\bibnamefont
  {Bouvard}}, \bibinfo {author} {\bibfnamefont {W.}~\bibnamefont {Herreman}},\
  and\ \bibinfo {author} {\bibfnamefont {F.}~\bibnamefont {Moisy}},\
  }\href@noop {} {\bibfield  {journal} {\bibinfo  {journal} {Phys. Rev.
  Fluids}\ }\textbf {\bibinfo {volume} {2}},\ \bibinfo {pages} {084801}
  (\bibinfo {year} {2017})}\BibitemShut {NoStop}%
\end{thebibliography}%


\begin{thebibliography}{26}%
\makeatletter
\providecommand \@ifxundefined [1]{%
 \@ifx{#1\undefined}
}%
\providecommand \@ifnum [1]{%
 \ifnum #1\expandafter \@firstoftwo
 \else \expandafter \@secondoftwo
 \fi
}%
\providecommand \@ifx [1]{%
 \ifx #1\expandafter \@firstoftwo
 \else \expandafter \@secondoftwo
 \fi
}%
\providecommand \natexlab [1]{#1}%
\providecommand \enquote  [1]{``#1''}%
\providecommand \bibnamefont  [1]{#1}%
\providecommand \bibfnamefont [1]{#1}%
\providecommand \citenamefont [1]{#1}%
\providecommand \href@noop [0]{\@secondoftwo}%
\providecommand \href [0]{\begingroup \@sanitize@url \@href}%
\providecommand \@href[1]{\@@startlink{#1}\@@href}%
\providecommand \@@href[1]{\endgroup#1\@@endlink}%
\providecommand \@sanitize@url [0]{\catcode `\\12\catcode `\$12\catcode
  `\&12\catcode `\#12\catcode `\^12\catcode `\_12\catcode `\%12\relax}%
\providecommand \@@startlink[1]{}%
\providecommand \@@endlink[0]{}%
\providecommand \url  [0]{\begingroup\@sanitize@url \@url }%
\providecommand \@url [1]{\endgroup\@href {#1}{\urlprefix }}%
\providecommand \urlprefix  [0]{URL }%
\providecommand \Eprint [0]{\href }%
\providecommand \doibase [0]{https://doi.org/}%
\providecommand \selectlanguage [0]{\@gobble}%
\providecommand \bibinfo  [0]{\@secondoftwo}%
\providecommand \bibfield  [0]{\@secondoftwo}%
\providecommand \translation [1]{[#1]}%
\providecommand \BibitemOpen [0]{}%
\providecommand \bibitemStop [0]{}%
\providecommand \bibitemNoStop [0]{.\EOS\space}%
\providecommand \EOS [0]{\spacefactor3000\relax}%
\providecommand \BibitemShut  [1]{\csname bibitem#1\endcsname}%
\let\auto@bib@innerbib\@empty
\bibitem [{\citenamefont {Faltinsen}\ and\ \citenamefont
  {Timokha}(2009)}]{faltinsen2005liquid}%
  \BibitemOpen
  \bibfield  {author} {\bibinfo {author} {\bibfnamefont {O.~M.}\ \bibnamefont
  {Faltinsen}}\ and\ \bibinfo {author} {\bibfnamefont {A.~N.}\ \bibnamefont
  {Timokha}},\ }\href@noop {} {\emph {\bibinfo {title} {Sloshing}}}\ (\bibinfo
  {publisher} {Cambridge University Press},\ \bibinfo {year}
  {2009})\BibitemShut {NoStop}%
\bibitem [{\citenamefont {Ibrahim}(2005)}]{ibrahim2009liquid}%
  \BibitemOpen
  \bibfield  {author} {\bibinfo {author} {\bibfnamefont {R.~A.}\ \bibnamefont
  {Ibrahim}},\ }\href@noop {} {\emph {\bibinfo {title} {Liquid sloshing
  dynamics: Theory and Applications}}}\ (\bibinfo  {publisher} {Cambridge
  University Press},\ \bibinfo {year} {2005})\BibitemShut {NoStop}%
\bibitem [{\citenamefont {Faltinsen}\ \emph {et~al.}(2016)\citenamefont
  {Faltinsen}, \citenamefont {Lukovsky},\ and\ \citenamefont
  {Timokha}}]{faltinsen2016resonant}%
  \BibitemOpen
  \bibfield  {author} {\bibinfo {author} {\bibfnamefont {O.~M.}\ \bibnamefont
  {Faltinsen}}, \bibinfo {author} {\bibfnamefont {I.~A.}\ \bibnamefont
  {Lukovsky}},\ and\ \bibinfo {author} {\bibfnamefont {A.~N.}\ \bibnamefont
  {Timokha}},\ }\href@noop {} {\bibfield  {journal} {\bibinfo  {journal}
  {J.~Fluid Mech.}\ }\textbf {\bibinfo {volume} {804}},\ \bibinfo {pages} {608}
  (\bibinfo {year} {2016})}\BibitemShut {NoStop}%
\bibitem [{\citenamefont {Marcotte}\ \emph {et~al.}(2023)\citenamefont
  {Marcotte}, \citenamefont {Gallaire},\ and\ \citenamefont
  {Bongarzone}}]{marcotte2023super}%
  \BibitemOpen
  \bibfield  {author} {\bibinfo {author} {\bibfnamefont {A.}~\bibnamefont
  {Marcotte}}, \bibinfo {author} {\bibfnamefont {F.}~\bibnamefont {Gallaire}},\
  and\ \bibinfo {author} {\bibfnamefont {A.}~\bibnamefont {Bongarzone}},\
  }\href@noop {} {\bibfield  {journal} {\bibinfo  {journal} {arXiv preprint
  arXiv:2302.02443}\ } (\bibinfo {year} {2023})}\BibitemShut {NoStop}%
\bibitem [{\citenamefont {Nayfeh}(2008)}]{nayfeh2008perturbation}%
  \BibitemOpen
  \bibfield  {author} {\bibinfo {author} {\bibfnamefont {A.~H.}\ \bibnamefont
  {Nayfeh}},\ }\href@noop {} {\emph {\bibinfo {title} {Perturbation methods}}}\
  (\bibinfo  {publisher} {John Wiley \& Sons},\ \bibinfo {year}
  {2008})\BibitemShut {NoStop}%
\bibitem [{\citenamefont {Bongarzone}\ \emph
  {et~al.}(2022{\natexlab{a}})\citenamefont {Bongarzone}, \citenamefont
  {Guido},\ and\ \citenamefont {Gallaire}}]{bongarzone2022amplitude}%
  \BibitemOpen
  \bibfield  {author} {\bibinfo {author} {\bibfnamefont {A.}~\bibnamefont
  {Bongarzone}}, \bibinfo {author} {\bibfnamefont {M.}~\bibnamefont {Guido}},\
  and\ \bibinfo {author} {\bibfnamefont {F.}~\bibnamefont {Gallaire}},\
  }\href@noop {} {\bibfield  {journal} {\bibinfo  {journal} {J.~Fluid Mech.}\
  }\textbf {\bibinfo {volume} {943}},\ \bibinfo {pages} {A28} (\bibinfo {year}
  {2022}{\natexlab{a}})}\BibitemShut {NoStop}%
\bibitem [{\citenamefont {Viola}\ and\ \citenamefont
  {Gallaire}(2018)}]{Viola2018b}%
  \BibitemOpen
  \bibfield  {author} {\bibinfo {author} {\bibfnamefont {F.}~\bibnamefont
  {Viola}}\ and\ \bibinfo {author} {\bibfnamefont {F.}~\bibnamefont
  {Gallaire}},\ }\href@noop {} {\bibfield  {journal} {\bibinfo  {journal}
  {Phys. Rev. Fluids}\ }\textbf {\bibinfo {volume} {3}},\ \bibinfo {pages}
  {094801} (\bibinfo {year} {2018})}\BibitemShut {NoStop}%
\bibitem [{\citenamefont {Bongarzone}\ \emph
  {et~al.}(2021{\natexlab{a}})\citenamefont {Bongarzone}, \citenamefont
  {Viola},\ and\ \citenamefont {Gallaire}}]{bongarzone2021relaxation}%
  \BibitemOpen
  \bibfield  {author} {\bibinfo {author} {\bibfnamefont {A.}~\bibnamefont
  {Bongarzone}}, \bibinfo {author} {\bibfnamefont {F.}~\bibnamefont {Viola}},\
  and\ \bibinfo {author} {\bibfnamefont {F.}~\bibnamefont {Gallaire}},\
  }\href@noop {} {\bibfield  {journal} {\bibinfo  {journal} {Chaos}\ }\textbf
  {\bibinfo {volume} {31}},\ \bibinfo {pages} {123124} (\bibinfo {year}
  {2021}{\natexlab{a}})}\BibitemShut {NoStop}%
\bibitem [{\citenamefont {Friedrichs}(2012)}]{friedrichs2012spectral}%
  \BibitemOpen
  \bibfield  {author} {\bibinfo {author} {\bibfnamefont {K.~O.}\ \bibnamefont
  {Friedrichs}},\ }\href@noop {} {\emph {\bibinfo {title} {Spectral theory of
  operators in Hilbert space}}}\ (\bibinfo  {publisher} {Springer Science \&
  Business Media},\ \bibinfo {year} {2012})\BibitemShut {NoStop}%
\bibitem [{\citenamefont {Bongarzone}\ \emph
  {et~al.}(2021{\natexlab{b}})\citenamefont {Bongarzone}, \citenamefont
  {Bertsch}, \citenamefont {Renaud},\ and\ \citenamefont
  {Gallaire}}]{bongarzone2021impinging}%
  \BibitemOpen
  \bibfield  {author} {\bibinfo {author} {\bibfnamefont {A.}~\bibnamefont
  {Bongarzone}}, \bibinfo {author} {\bibfnamefont {A.}~\bibnamefont {Bertsch}},
  \bibinfo {author} {\bibfnamefont {P.}~\bibnamefont {Renaud}},\ and\ \bibinfo
  {author} {\bibfnamefont {F.}~\bibnamefont {Gallaire}},\ }\href@noop {}
  {\bibfield  {journal} {\bibinfo  {journal} {J.~Fluid Mech.}\ }\textbf
  {\bibinfo {volume} {913}} (\bibinfo {year} {2021}{\natexlab{b}})}\BibitemShut
  {NoStop}%
\bibitem [{\citenamefont {Bongarzone}\ \emph
  {et~al.}(2022{\natexlab{b}})\citenamefont {Bongarzone}, \citenamefont
  {Viola}, \citenamefont {Camarri},\ and\ \citenamefont
  {Gallaire}}]{bongarzone2022sub}%
  \BibitemOpen
  \bibfield  {author} {\bibinfo {author} {\bibfnamefont {A.}~\bibnamefont
  {Bongarzone}}, \bibinfo {author} {\bibfnamefont {F.}~\bibnamefont {Viola}},
  \bibinfo {author} {\bibfnamefont {S.}~\bibnamefont {Camarri}},\ and\ \bibinfo
  {author} {\bibfnamefont {F.}~\bibnamefont {Gallaire}},\ }\href@noop {}
  {\bibfield  {journal} {\bibinfo  {journal} {J.~Fluid Mech.}\ }\textbf
  {\bibinfo {volume} {947}},\ \bibinfo {pages} {A24} (\bibinfo {year}
  {2022}{\natexlab{b}})}\BibitemShut {NoStop}%
\bibitem [{\citenamefont {Viola}\ \emph {et~al.}(2018)\citenamefont {Viola},
  \citenamefont {Brun},\ and\ \citenamefont {Gallaire}}]{Viola2018a}%
  \BibitemOpen
  \bibfield  {author} {\bibinfo {author} {\bibfnamefont {F.}~\bibnamefont
  {Viola}}, \bibinfo {author} {\bibfnamefont {P.-T.}\ \bibnamefont {Brun}},\
  and\ \bibinfo {author} {\bibfnamefont {F.}~\bibnamefont {Gallaire}},\
  }\href@noop {} {\bibfield  {journal} {\bibinfo  {journal} {J.~Fluid Mech.}\
  }\textbf {\bibinfo {volume} {837}},\ \bibinfo {pages} {788} (\bibinfo {year}
  {2018})}\BibitemShut {NoStop}%
\bibitem [{\citenamefont {Raynovskyy}\ and\ \citenamefont
  {Timokha}(2018)}]{raynovskyy2018damped}%
  \BibitemOpen
  \bibfield  {author} {\bibinfo {author} {\bibfnamefont {I.~A.}\ \bibnamefont
  {Raynovskyy}}\ and\ \bibinfo {author} {\bibfnamefont {A.~N.}\ \bibnamefont
  {Timokha}},\ }\href@noop {} {\bibfield  {journal} {\bibinfo  {journal} {Fluid
  Dyn. Res.}\ }\textbf {\bibinfo {volume} {50}},\ \bibinfo {pages} {045502}
  (\bibinfo {year} {2018})}\BibitemShut {NoStop}%
\bibitem [{\citenamefont {Lamb}(1993)}]{Lamb32}%
  \BibitemOpen
  \bibfield  {author} {\bibinfo {author} {\bibfnamefont {H.}~\bibnamefont
  {Lamb}},\ }\href@noop {} {\emph {\bibinfo {title} {Hydrodynamics}}}\
  (\bibinfo  {publisher} {Cambridge University Press},\ \bibinfo {year}
  {1993})\BibitemShut {NoStop}%
\bibitem [{\citenamefont {Cocciaro}\ \emph {et~al.}(1993)\citenamefont
  {Cocciaro}, \citenamefont {Faetti},\ and\ \citenamefont
  {Festa}}]{Cocciaro93}%
  \BibitemOpen
  \bibfield  {author} {\bibinfo {author} {\bibfnamefont {B.}~\bibnamefont
  {Cocciaro}}, \bibinfo {author} {\bibfnamefont {S.}~\bibnamefont {Faetti}},\
  and\ \bibinfo {author} {\bibfnamefont {C.}~\bibnamefont {Festa}},\
  }\href@noop {} {\bibfield  {journal} {\bibinfo  {journal} {J.~Fluid Mech.}\
  }\textbf {\bibinfo {volume} {246}},\ \bibinfo {pages} {43} (\bibinfo {year}
  {1993})}\BibitemShut {NoStop}%
\bibitem [{\citenamefont {Case}\ and\ \citenamefont
  {Parkinson}(1957)}]{Case1957}%
  \BibitemOpen
  \bibfield  {author} {\bibinfo {author} {\bibfnamefont {K.~M.}\ \bibnamefont
  {Case}}\ and\ \bibinfo {author} {\bibfnamefont {W.~C.}\ \bibnamefont
  {Parkinson}},\ }\href@noop {} {\bibfield  {journal} {\bibinfo  {journal}
  {J.~Fluid Mech.}\ }\textbf {\bibinfo {volume} {2}},\ \bibinfo {pages} {172}
  (\bibinfo {year} {1957})}\BibitemShut {NoStop}%
\bibitem [{\citenamefont {Henderson}\ and\ \citenamefont
  {Miles}(1994)}]{henderson1994surface}%
  \BibitemOpen
  \bibfield  {author} {\bibinfo {author} {\bibfnamefont {D.~M.}\ \bibnamefont
  {Henderson}}\ and\ \bibinfo {author} {\bibfnamefont {J.~W.}\ \bibnamefont
  {Miles}},\ }\href@noop {} {\bibfield  {journal} {\bibinfo  {journal}
  {J.~Fluid Mech.}\ }\textbf {\bibinfo {volume} {275}},\ \bibinfo {pages} {285}
  (\bibinfo {year} {1994})}\BibitemShut {NoStop}%
\bibitem [{\citenamefont {Keulegan}(1959)}]{Keulegan59}%
  \BibitemOpen
  \bibfield  {author} {\bibinfo {author} {\bibfnamefont {G.~H.}\ \bibnamefont
  {Keulegan}},\ }\href@noop {} {\bibfield  {journal} {\bibinfo  {journal}
  {J.~Fluid Mech.}\ }\textbf {\bibinfo {volume} {6}},\ \bibinfo {pages} {33}
  (\bibinfo {year} {1959})}\BibitemShut {NoStop}%
\bibitem [{\citenamefont {Dussan}(1979)}]{Dussan79}%
  \BibitemOpen
  \bibfield  {author} {\bibinfo {author} {\bibfnamefont {E.~B.}\ \bibnamefont
  {Dussan}},\ }\href@noop {} {\bibfield  {journal} {\bibinfo  {journal} {Annu.
  Rev. Fluid Mech.}\ }\textbf {\bibinfo {volume} {11}},\ \bibinfo {pages} {371}
  (\bibinfo {year} {1979})}\BibitemShut {NoStop}%
\bibitem [{\citenamefont {Hocking}(1987)}]{Hocking87}%
  \BibitemOpen
  \bibfield  {author} {\bibinfo {author} {\bibfnamefont {L.~M.}\ \bibnamefont
  {Hocking}},\ }\href@noop {} {\bibfield  {journal} {\bibinfo  {journal}
  {J.~Fluid Mech.}\ }\textbf {\bibinfo {volume} {179}},\ \bibinfo {pages} {253}
  (\bibinfo {year} {1987})}\BibitemShut {NoStop}%
\bibitem [{\citenamefont {Dollet}\ \emph {et~al.}(2020)\citenamefont {Dollet},
  \citenamefont {Lorenceau},\ and\ \citenamefont
  {Gallaire}}]{dollet2020transition}%
  \BibitemOpen
  \bibfield  {author} {\bibinfo {author} {\bibfnamefont {B.}~\bibnamefont
  {Dollet}}, \bibinfo {author} {\bibfnamefont {{\'E}.}~\bibnamefont
  {Lorenceau}},\ and\ \bibinfo {author} {\bibfnamefont {F.}~\bibnamefont
  {Gallaire}},\ }\href@noop {} {\bibfield  {journal} {\bibinfo  {journal}
  {Phys. Rev. Lett.}\ }\textbf {\bibinfo {volume} {124}},\ \bibinfo {pages}
  {104502} (\bibinfo {year} {2020})}\BibitemShut {NoStop}%
\bibitem [{\citenamefont {Raynovskyy}\ and\ \citenamefont
  {Timokha}(2020)}]{raynovskyy2020sloshing}%
  \BibitemOpen
  \bibfield  {author} {\bibinfo {author} {\bibfnamefont {I.~A.}\ \bibnamefont
  {Raynovskyy}}\ and\ \bibinfo {author} {\bibfnamefont {A.~N.}\ \bibnamefont
  {Timokha}},\ }\href@noop {} {\emph {\bibinfo {title} {Sloshing in Upright
  Circular Containers: Theory, Analytical Solutions, and Applications}}}\
  (\bibinfo  {publisher} {CRC Press},\ \bibinfo {year} {2020})\BibitemShut
  {NoStop}%
\bibitem [{\citenamefont {Hutton}(1964)}]{hutton1964fluid}%
  \BibitemOpen
  \bibfield  {author} {\bibinfo {author} {\bibfnamefont {R.~E.}\ \bibnamefont
  {Hutton}},\ }\href@noop {} {\bibfield  {journal} {\bibinfo  {journal} {Trans.
  ASME J. Appl. Mech.}\ }\textbf {\bibinfo {volume} {31}},\ \bibinfo {pages}
  {145} (\bibinfo {year} {1964})}\BibitemShut {NoStop}%
\bibitem [{\citenamefont {Faltinsen}\ and\ \citenamefont
  {Timokha}(2019)}]{faltinsen2019inviscid}%
  \BibitemOpen
  \bibfield  {author} {\bibinfo {author} {\bibfnamefont {O.~M.}\ \bibnamefont
  {Faltinsen}}\ and\ \bibinfo {author} {\bibfnamefont {A.~N.}\ \bibnamefont
  {Timokha}},\ }\href@noop {} {\bibfield  {journal} {\bibinfo  {journal}
  {J.~Fluid Mech.}\ }\textbf {\bibinfo {volume} {865}},\ \bibinfo {pages} {884}
  (\bibinfo {year} {2019})}\BibitemShut {NoStop}%
\bibitem [{\citenamefont {B{\"a}uerlein}\ and\ \citenamefont
  {Avila}(2021)}]{bauerlein2021phase}%
  \BibitemOpen
  \bibfield  {author} {\bibinfo {author} {\bibfnamefont {B.}~\bibnamefont
  {B{\"a}uerlein}}\ and\ \bibinfo {author} {\bibfnamefont {K.}~\bibnamefont
  {Avila}},\ }\href@noop {} {\bibfield  {journal} {\bibinfo  {journal}
  {J.~Fluid Mech.}\ }\textbf {\bibinfo {volume} {925}} (\bibinfo {year}
  {2021})}\BibitemShut {NoStop}%
\bibitem [{\citenamefont {Reclari}\ \emph {et~al.}(2014)\citenamefont
  {Reclari}, \citenamefont {Dreyer}, \citenamefont {Tissot}, \citenamefont
  {Obreschkow}, \citenamefont {Wurm},\ and\ \citenamefont
  {Farhat}}]{reclari2014surface}%
  \BibitemOpen
  \bibfield  {author} {\bibinfo {author} {\bibfnamefont {M.}~\bibnamefont
  {Reclari}}, \bibinfo {author} {\bibfnamefont {M.}~\bibnamefont {Dreyer}},
  \bibinfo {author} {\bibfnamefont {S.}~\bibnamefont {Tissot}}, \bibinfo
  {author} {\bibfnamefont {D.}~\bibnamefont {Obreschkow}}, \bibinfo {author}
  {\bibfnamefont {F.~M.}\ \bibnamefont {Wurm}},\ and\ \bibinfo {author}
  {\bibfnamefont {M.}~\bibnamefont {Farhat}},\ }\href@noop {} {\bibfield
  {journal} {\bibinfo  {journal} {Phys. Fluids}\ }\textbf {\bibinfo {volume}
  {26}},\ \bibinfo {pages} {052104} (\bibinfo {year} {2014})}\BibitemShut
  {NoStop}%
\end{thebibliography}%

\end{document}


\preprint{APS/123-QED}

\title{Swirling against the forcing: Evidence of stable counter-directed sloshing waves in orbital-shaken reservoirs\\ Supplementary Material}

\author{Alice Marcotte, François Gallaire}
 \email{francois.gallaire@epfl.ch}
 \author{Alessandro Bongarzone}
\affiliation{%
 Laboratory of Fluid Mechanics and Instabilities, École Polytechnique Fédérale de Lausanne, Lausanne, CH-1015, Switzerland
}%

\date{\today}

\begin{abstract}
In this supplementary document, we provide further details about the derivation of the asymptotic model, the estimation of the heuristic damping coefficient and the numerical computation of the stable wave amplitudes.
\end{abstract}

\maketitle


\section{Governing equations}\label{sec:Sec1}

In the moving reference frame, any planar elliptic-like shaking can be represented by the following equations describing the motion acceleration of the container axis intersection with the $z=0$ plane (located at the unperturbed free surface), parametrized in polar coordinates ($r$, $\theta$),
\begin{equation}
\label{eq:EqMotWall_ELL}
\frac{\text{d}^2\mathbf{X}_0}{\text{d}t^2}=
  \begin{cases}
    \left(-f_x\cos{\left(\Omega t\right)}\cos\theta-f_y\sin{\left(\Omega t\right)}\sin\theta\right)\,\mathbf{e}_r,\\
   \left(\, \, \, \, f_x\cos{\left(\Omega t\right)}\sin\theta-f_y\sin{\left(\Omega t\right)}\sin\theta\right)\,\mathbf{e}_{\theta},
  \end{cases}
\end{equation}
\noindent with $f_x=\overline{a}_x\Omega^2/R$ and $f_y=\overline{a}_y\Omega^2/R$ are the non-dimensional major- and minor-axis driving acceleration components, respectively, and $\Omega=\overline{\Omega}/\sqrt{g/R}$ the non-dimensional driving angular frequency. The bar symbol refers to the dimensional quantities.\\
\indent In the potential flow limit, i.e. the flow is assumed inviscid, irrotational and incompressible, the liquid motion is governed by the Laplace equation, subjected to the homogeneous no-penetration condition at the solid lateral wall, $r=R$, and bottom $z=-h$,
\begin{equation}
\label{eq:GovEq_Lap}
\Delta\Phi=0,\ \ \ \ \nabla\Phi\cdot\mathbf{n}=\mathbf{0},
\end{equation}
and by the kinematic and dynamic boundary conditions at the free surface $z=\eta\left(r,\theta,t\right)$ \citep{faltinsen2005liquid,ibrahim2009liquid},
\begin{subequations}
\begin{equation}
\label{eq:GovEq_Kin}
\frac{\partial \eta}{\partial t}+\nabla\Phi\cdot\nabla\eta-\frac{\partial\Phi}{\partial z}=0.
\end{equation}
\begin{eqnarray}
\label{eq:GovEq_Dyn}
\frac{\partial\Phi}{\partial t}+\frac{1}{2}\nabla\Phi\cdot\nabla\Phi+\eta=\\
=r\left(f_x\cos{\left(\Omega t\right)}\cos{\theta}+f_y\sin{\left(\Omega t\right)}\sin{\theta}\right),\notag
\end{eqnarray}
\end{subequations}
\noindent made non-dimensional using the characteristic length $R$ and velocity $\sqrt{gR}$. $\Phi\left(r,\theta,z,t\right)$ and $\eta\left(r,\theta,t\right)$ denote potential velocity field and free surface elevation, respectively. Note that, as in \cite{faltinsen2016resonant} and \cite{marcotte2023super}, surface tension effects have been neglected. By introducing the orbit aspect ratio $\alpha=\overline{a}_y/\overline{a}_x=f_y/f_x$, so that $f_x=f$ and $f_y=\alpha f$, equation~\eqref{eq:GovEq_Dyn} can be conveniently rewritten as\\
\begin{eqnarray}
\label{eq:GovEq_Dyn_new}
\frac{\partial\Phi}{\partial t}+\frac{1}{2}\nabla\Phi\cdot\nabla\Phi+\eta=\\
r\frac{f}{2}\left(\alpha_{_A}e^{\text{i}\left(\Omega t-\theta\right)}+\alpha_{_B}e^{\text{i}\left(\Omega t+\theta\right)}\right)+c.c.\,,\notag
\end{eqnarray}
\noindent with $c.c.$ denoting the complex conjugate and with $\alpha_{_A}=\left(1+\alpha\right)/2$ and $\alpha_{_B}=\left(1-\alpha\right)/2$ two auxiliary orbit parameter (see Appendix~B of \cite{marcotte2023super}).

\section{Multiple timescale weakly nonlinear model}\label{sec:Sec1bis}

\begin{figure}
\includegraphics[width=0.44\textwidth]{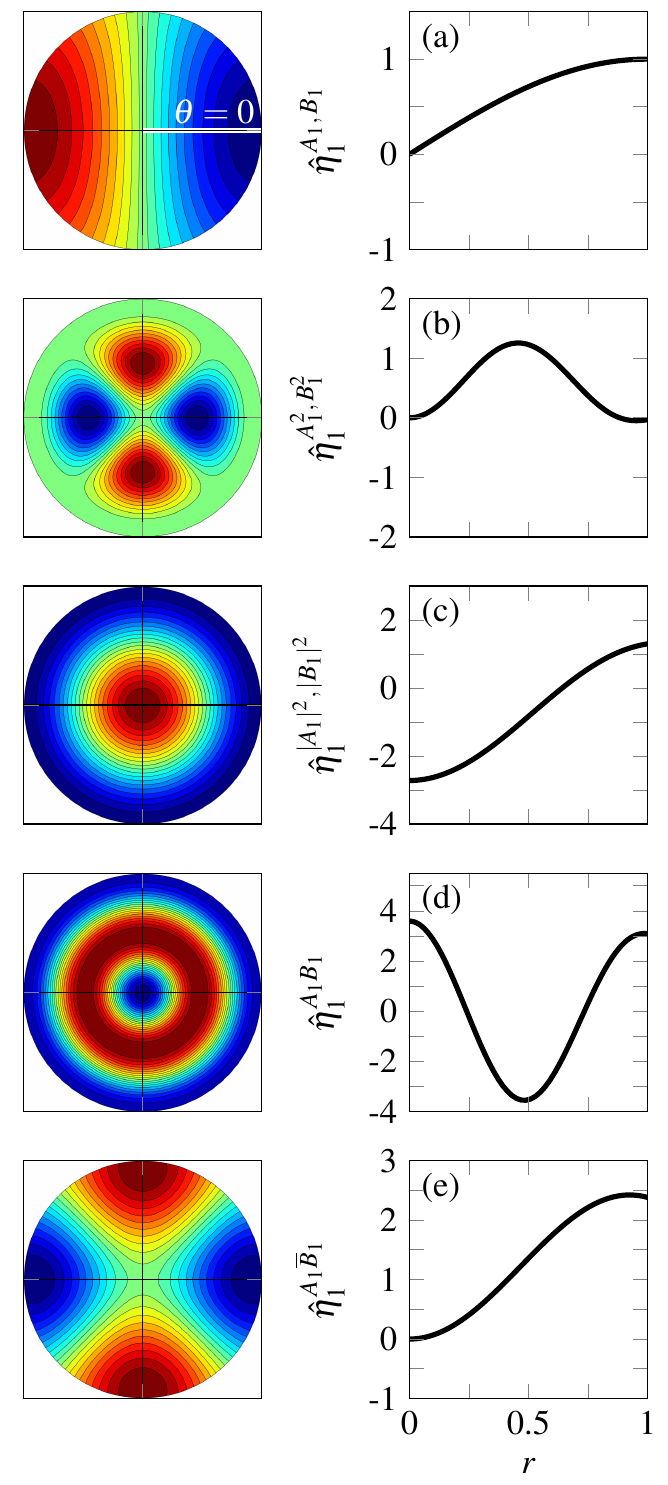}
\caption{(a) First-order and (b)-(e) second-order free surface deformations. \textit{Right}, interface as a function of the radial coordinate only and at $\theta=0$, e.g. $\hat{\eta}_1^{A1,\,B_1}\left(r\right)$. \textit{Left}, top-view of the full surface deformations, rebuilt according to the corresponding azimuthal periodicity and shown for $t=0$, i.e. $\hat{\eta}_1^{A_1,\,B_1}\left(r\right)\cos{m\theta}$. The first-order solution is normalized with the amplitude and phase of the contact line elevation (at $r=1$), such that the free surface $\eta_1^{A_1,\,B_1}$ is purely real, whereas the potential velocity field $\hat{\Phi}_1^{A_1,\,B_1}$ is purely imaginary. Note that, owing to the symmetries of the problem, the system admits the following invariant transformation: $\left(\hat{\mathbf{q}},+m,\text{i}\omega_0\right)\longrightarrow\left(\hat{\mathbf{q}},-m,\text{i}\omega_0\right)$, so that $\hat{\eta}_1^{A_1}=\hat{\eta}_1^{B_1}$, $\hat{\eta}_2^{|A_1|^2}=\hat{\eta}_2^{|B_1|^2}$ and $\hat{\eta}_2^{A_1^2}=\hat{\eta}_2^{B_1^2}$. In other words, only part of the first and second order responses need to be computed explicitly.}
\label{fig:Fig1sm} 
\end{figure}

The weakly nonlinear multiple timescale analysis formalized in \S4 of \cite{marcotte2023super} is based on the following asymptotic expansion for the flow quantities,
\begin{equation}
    \label{eq:asy_exp}
    \mathbf{q}\left(r,\theta,z,t\right)=\left\{\Phi,\eta\right\}^T=\epsilon\mathbf{q}_1+\epsilon^2\mathbf{q}_2+\epsilon^3\mathbf{q}_3+\text{O}\left(\epsilon^4\right),
\end{equation}
\noindent and on the assumption of a small forcing amplitude of order $f=\epsilon^3F$. We also allow for a small frequency detuning with respect to the first system's natural frequency, $\omega_0$, such that $\Omega=\omega_0+\lambda$, with $\lambda=\epsilon^2\Lambda$ and $\epsilon$ a small parameter $\ll1$. The new auxiliary parameters $F$ and $\Lambda$ are assumed of order $\text{O}\left(1\right)$.\\
\indent We note that the forcing amplitude could be assumed of order $\epsilon^2$ (as the other parameters); however, this complicates unnecessarily the second-order problem without modifying the final amplitude equation, even if the values of its coefficients end up being slightly (up to order $\epsilon$) different.\\
\indent As suggested by the azimuthal periodicity of the forcing term on the right-hand-side of~\eqref{eq:GovEq_Dyn}, $m=\pm1$ (with $m$ a so-called azimuthal wavenumber), we assume a leading order solution as the sum of two counter-propagating traveling waves
\begin{eqnarray}
\label{eq:eps1}
\mathbf{q}_1\left(r,\theta,z,t\right)=A_1\left(T_2\right)\hat{\mathbf{q}}_{1}^{A_1}\left(r,z\right)e^{\text{i}\left(\omega_0 t-\theta\right)}\ \ \ \ \ \ \ \ \ \\
+B_1\left(T_2\right)\hat{\mathbf{q}}_{1}^{B_1}\left(r,z\right)e^{\text{i}\left(\omega_0 t+\theta\right)}+c.c.\,.\notag
\end{eqnarray}
\noindent As typical of multiple timescale analyses \citep{nayfeh2008perturbation}, the complex amplitudes $A_1$ and $B_1$, functions of the slow time scale $T_2=\epsilon^2 t$ and still undetermined at this stage of the expansion, describe the slow time amplitude modulation of the two oscillating waves and must be determined at a higher order of the asymptotic expansion.\\
\indent The natural frequency $\omega_0$ and structure $\hat{\mathbf{q}}_1^{A_1}$ (and $\hat{\mathbf{q}}_1^{B_1}$) assume the meaning of eigenvalue and associated eigenmode of the leading order linearized sloshing operator, whose matrix compact form can be written as $\left(\text{i}\omega_0\mathcal{B}-\mathcal{A}_{m=\pm1}\right)\hat{\mathbf{q}}_1^{A_1,\,B_1}=0$ (see \cite{marcotte2023super}). Despite the fact that an exact analytical solution to this generalized eigenvalue problem can be obtained via a Bessel-function representation, in this work, as in \cite{bongarzone2022amplitude}, we have opted for a numerical scheme based on a discretization technique, where operators $B$ and $A_{m}$ are discretized in space by means of a Gauss-Lobatto-Chebyshev pseudo-spectral collocation method with a two-dimensional mapping implemented in \textit{Matlab}, which is analogous to that described in \cite{Viola2018b} and \cite{bongarzone2021relaxation}.\\
\indent By pursuing the expansion to the second order in $\epsilon$, one obtains a linear system forced by second order non-linear terms produced by combinations of the two leading order waves through, e.g., $\nabla\Phi_1\cdot\nabla\Phi_1/2$ in the dynamic condition and $\nabla\Phi_1\cdot\nabla\eta_1$ in the kinematic equation. These forcing terms, $\boldsymbol{\mathcal{F}}_2$, are proportional to $A_1^2$ and $B_1^2$ (second harmonics), to $|A_1|^2$ and $|B_1|^2$ (steady and axisymmetric mean flow corrections) and to $A_1B_1$ and $A_1\overline{B}_1$ (cross-quadratic interactions), and therefore they call for a second-order solution in the form
\begin{eqnarray}
\label{eq:eps2}
\mathbf{q}_2=|A_1|^2\hat{\mathbf{q}}_2^{A_1\bar{A}_1}+|B_1|^2\hat{\mathbf{q}}_2^{B_1\bar{B}_1}\\
+\left(A_1^2\hat{\mathbf{q}}_{2}^{A_1A_1}e^{\text{i}2\left(\omega_0 t-\theta\right)}+B_1^2\hat{\mathbf{q}}_{2}^{B_1B_1}e^{\text{i}2\left(\omega_0 t+\theta\right)}+c.c.\right)\notag\\
+\left(A_1B_1\hat{\mathbf{q}}_{2}^{A_1B_1}e^{\text{i}2\omega_0 t}+A_1\overline{B}_1\hat{\mathbf{q}}_{2}^{A_1\overline{B}_1}e^{-\text{i}2\theta}+c.c.\right). \ \notag
\end{eqnarray}
\noindent None of the associated forcing terms being resonant, each spatial structure, function of the $\left(r,z\right)$-space, can be computed numerically as described in \cite{bongarzone2022amplitude} by simply inverting a linear operator. The resulting structures are shown in Fig.~\ref{fig:Fig1sm} in terms of free surface deformations.\\
We now move forward to the $\epsilon^3$--order problem, which is once again a linear problem forced by combinations of the first~\eqref{eq:eps1} and second order~\eqref{eq:eps2} solutions, produced by third order non-linearities through, e.g., $\left(\nabla\Phi_1\cdot\nabla\Phi_2+\nabla\Phi_2\cdot\nabla\Phi_1\right)/2$ in the dynamic condition or $\nabla\Phi_1\cdot\nabla\eta_2+\nabla\Phi_2\cdot\nabla\eta_1$ in the kinematic equation, as well as by the slow time-$T_2$ derivative of the leading order solution and by the external forcing, which was assumed of order $\epsilon^3$:
\begin{eqnarray}
\label{eq:SC_eps3}
\left(\partial_t\mathbf{B}-\textcolor{black}{\mathbf{A}_{m}}\right)\mathbf{q}_3=\boldsymbol{\mathcal{F}}_3=\ \ \ \ \\
=-\frac{\partial A_1}{\partial T_2} \mathbf{B}\hat{\mathbf{q}}_1^{A_1}e^{\text{i}\left(\omega_0 t-\theta\right)}-\frac{\partial B_1}{\partial T_2}\mathbf{B}\hat{\mathbf{q}}_1^{B_1}e^{\text{i}\left(\omega_0 t+\theta\right)}\notag \\
+ |A_1|^2A_1 \boldsymbol{\hat{\mathcal{F}}}_3^{|A_1|^2A_1}e^{\text{i}\left(\omega_0 t-\theta\right)}+ |B_1|^2B_1 \boldsymbol{\hat{\mathcal{F}}}_3^{|B_1|^2B_1}e^{\text{i}\left(\omega_0 t+\theta\right)}\notag \\
+ |B_1|^2A_1 \boldsymbol{\hat{\mathcal{F}}}_3^{|B_1|^2A_1}e^{\text{i}\left(\omega_0 t-\theta\right)}+ |A_1|^2B_1 \boldsymbol{\hat{\mathcal{F}}}_3^{|A_1|^2B_1}e^{\text{i}\left(\omega_0 t+\theta\right)}\notag \\
+\alpha_{_A}F\boldsymbol{\hat{\mathcal{F}}}_3^F e^{\text{i}\left(\omega_0 t-\theta\right)}e^{\text{i}\Lambda T_2}+\alpha_{_B}F\boldsymbol{\hat{\mathcal{F}}}_3^F e^{\text{i}\left(\omega_0 t+\theta\right)}e^{\text{i}\Lambda T_2}\notag \\
+\text{N.R.T.}+c.c.,\nonumber
\end{eqnarray}
with $\boldsymbol{\hat{\mathcal{F}}}_3^F=\left\{0,r/2\right\}^T$ and where $\text{N.R.T.}$ stands for non-resonating terms. The latter terms are not strictly relevant for further analysis and can therefore be neglected. \textcolor{black}{Amplitudes equations for $A_1$ and $B_1$ are obtained by requiring that secular terms do not appear in the solution to equation~\eqref{eq:SC_eps3}, where secularity} results from all resonant forcing terms in $\boldsymbol{\mathcal{F}}_3$ (see Appendix~D of \cite{bongarzone2022amplitude} for its explicit expression), i.e. all terms sharing the same frequency and wavenumber of $\mathbf{q}_1$, e.g. $\left(\omega_0,m=\pm1\right)$, and in effect all terms explicitly written in~\eqref{eq:SC_eps3}. It follows that a compatibility condition must be enforced through the Fredholm alternative \citep{friedrichs2012spectral}, which imposes the amplitudes $A=\epsilon A_1e^{-\text{i}\lambda t}$ and $B=\epsilon B_1e^{-\text{i}\lambda t}$ to obey the following normal form 
\begin{subequations}
\begin{equation}
\label{eq:AmpEqSCfinalA}
\frac{\text{d}A}{\text{d}t}=-\text{i}\lambda A + \text{i}\,\alpha_{_A}\mu f + \text{i}\,\nu |A|^2A +\text{i}\,\xi |B|^2A,
\end{equation}
\begin{equation}
\label{eq:AmpEqSCfinalB}
\frac{\text{d}B}{\text{d}t}=-\text{i}\lambda B + \text{i}\,\alpha_{_B}\mu f + \text{i}\,\nu |B|^2B + \text{i}\,\xi |A|^2B,
\end{equation}
\end{subequations}
where the physical time $t=T_2/\epsilon^2$ has been reintroduced and where forcing amplitude and detuning parameter are recast in terms of their corresponding physical values, $f=\epsilon^3 F$ and $\lambda=\epsilon^2\Lambda=\Omega-\omega_0$, so as to eliminate the small implicit parameter $\epsilon$ \citep{bongarzone2021impinging,bongarzone2022sub}. The various normal form coefficients, which turn out to be real-valued quantities due to the absence of dissipation, are computed as scalar products between the adjoint mode, $\left(\hat{\mathbf{q}}_1^{A_1 \dagger},\hat{\mathbf{q}}_1^{B_1 \dagger}\right)$, associated with $\left(\hat{\mathbf{q}}_1^{A_1},\hat{\mathbf{q}}_1^{B_1}\right)$, and the third order resonant forcing terms:\\
\begin{subequations}
\begin{eqnarray}
\label{eq:SC_eps3_coeff_mu}
\text{i}\,\mathcal{I}_{_{SC}}\,\mu=<\hat{\mathbf{q}}_1^{A_1 \dagger},\boldsymbol{\hat{\mathcal{F}}}_3^F>=\int_{0}^{1}\left(r/2\right)\overline{\hat{\eta}}_1^{A_1 \dagger}\, r\text{d}r,
\end{eqnarray}
\begin{eqnarray}
\label{eq:SC_eps3_coeff_nu}
\text{i}\,\mathcal{I}_{_{SC}}\,\nu=<\hat{\mathbf{q}}_1^{A_1 \dagger},\boldsymbol{\hat{\mathcal{F}}}_3^{|A_1|^2A_1}>=\\
=\int_{0}^{1}\left(\overline{\hat{\eta}}_1^{A_1 \dagger}\hat{\mathcal{F}}_{3_{\text{dyn}}}^{|A_1|^2A_1}+\overline{\hat{\Phi}}_1^{A_1 \dagger}\hat{\mathcal{F}}_{3_{\text{kin}}}^{|A_1|^2A_1}\right)\, r\text{d}r,\notag
\end{eqnarray}
\begin{eqnarray}
\label{eq:SC_eps3_coeff_xi}
\text{i}\,\mathcal{I}_{_{SC}}\,\xi=<\hat{\mathbf{q}}_1^{A_1 \dagger},\boldsymbol{\hat{\mathcal{F}}}_3^{|B_1|^2A_1}>=\\
=\int_{0}^{1}\left(\overline{\hat{\eta}}_1^{A_1 \dagger}\hat{\mathcal{F}}_{3_{\text{dyn}}}^{|B_1|^2A_1}+\overline{\hat{\Phi}}_1^{A_1 \dagger}\hat{\mathcal{F}}_{3_{\text{kin}}}^{|B_1|^2A_1}\right)\, r\text{d}r.\notag
\end{eqnarray}
\end{subequations}
\noindent where 
\begin{eqnarray}
\mathcal{I}_{_{SC}}=<\hat{\mathbf{q}}_1^{A_1 \dagger},\mathbf{B}\hat{\mathbf{q}}_1^{A_1}>=\\
=\int_{0}^{1}\left(\overline{\hat{\eta}}_1^{A_1 \dagger}\hat{\Phi}_1^{A_1}+\overline{\hat{\Phi}}_1^{A_1 \dagger}\hat{\eta}_1^{A_1}\right)\, r\text{d}r.\notag
\end{eqnarray}
\noindent Here $\left(\hat{\mathbf{q}}_1^{A_1 \dagger},\hat{\mathbf{q}}_1^{B_1 \dagger}\right)=\left(\overline{\hat{\mathbf{q}}}_1^{A_1},\overline{\hat{\mathbf{q}}}_1^{B_1}\right)$, since the inviscid problem is self--adjoint with respect to the Hermitian scalar product $<\mathbf{a},\mathbf{b}>=\int_{V}^{}\overline{\mathbf{a}}\cdot\mathbf{b}\,\text{d}V$, with $\mathbf{a}$ and $\mathbf{b}$ two generic vector (see \cite{Viola2018a} for a thorough discussion and derivation of the adjoint problem).\\
\indent For the sake of brevity, we do not report the expression of the various forcing terms. As an example, the full expression of $\hat{\mathcal{F}}_{3_{kin}}^{|A_1|^2A_1}$ is given in Appendix~D of \cite{bongarzone2022amplitude}. The other forcing terms are calculated analogously.

\section{Phenomenological damping coefficient}

In contradistinction with the original system proposed in Appendix~B~(B$\,5a$)-(B$\,5b$) of \cite{marcotte2023super}, we account here for a heuristic damping coefficient, $\sigma$, whose value, by analogy with \cite{raynovskyy2018damped}, is estimated according to the well-known expression
 \begin{eqnarray}
\label{eq:dampM}
\sigma=\underbrace{\frac{2k^2}{Re}}_{\text{bulk}}+\sqrt{\frac{\omega_0}{2Re}}\left(\underbrace{\frac{k\cosh^2{kH}}{\sinh{2kH}}}_{\text{surf. contamination}}\right)\ \ \ \ \ \ \\
+\sqrt{\frac{\omega_0}{2Re}}\left(\underbrace{\frac{k}{\sinh{2kH}}}_{\text{bottom}}+\underbrace{\frac{1}{2}\frac{1+\left(1/k\right)^2}{1-\left(1/k\right)^2}-\frac{kH}{\sinh{2kH}}}_{\textit{sidewall}}\right).\notag
\end{eqnarray}
\noindent The damping associated with lowest natural frequency, $\omega_0=\overline{\omega}_0/\sqrt{g/R}=\sqrt{k\tanh{\left(kH\right)}}=1.3547$ (with wavenumber $k=1.8412$) \citep{Lamb32}, in a container of diameter $D=2R=0.172\,\text{m}$ filled to a depth $H=h/R=1.744$ with distilled water, i.e. $\rho=1000\,\text{kg/m$^3$}$, $\mu=0.001\,\text{kg/ms}$ and $\gamma=0.072\,\text{N/m}$, for which $Re=\rho\sqrt{g R^3}/\mu=78\,952$ (Reynolds number), amounts to $\sigma=0.0055$.\\
\indent Typically the viscous damping rate can be interpreter as a slow damping process \citep{Cocciaro93,Viola2018b}, i.e. $1/\sigma\approx180$, over a faster time scale represented by the wave oscillation, i.e. $1/\omega_0\approx0.5$. When this hypothesis holds, as in the present experimental study, the damping coefficient is assumed to be small of order $\epsilon^2$, such that damping terms as $\sigma A$ and $\sigma B$, both of order~$\sim\text{O}\left(\epsilon^3\right)$ ($A,B\sim\text{O}\left(\epsilon\right)$), can be phenomenologically added \textit{a posteriori} to the final inviscid amplitude equations.\\
\indent Expression~\eqref{eq:dampM} englobes different effects, i.e. viscous dissipation occurring in the Stokes boundary layers (at the solid lateral and bottom wall), bulk dissipation and possible sources of dissipation associated with free surface contamination effects \citep{Case1957,henderson1994surface}. However, it does not account for any form of dissipation induced by contact angle dynamics \citep{Keulegan59,Dussan79,Hocking87,dollet2020transition,bongarzone2021relaxation}, by wave breaking and overtuning \citep{raynovskyy2020sloshing} or by Prandtl mass-transport phenomena \citep{hutton1964fluid,faltinsen2019inviscid}.\\
\indent Moreover, as pointed out in Appendix~A of \cite{bongarzone2022amplitude}, prediction~\eqref{eq:dampM} is only valid for free capillary-gravity waves, whereas the dissipation rates of forced wave motions is generally more complex, i.e. it is typically a function of the wave amplitude \citep{raynovskyy2020sloshing}. A rigorous viscous analysis would indeed produce complex eigenfunctions and, therefore, complex-valued normal form coefficients \citep{bongarzone2022sub}, e.g. $\nu_{_{SC}}=\text{Re}\left[\nu_{_{SC}}\right]+\text{i}\,\text{Im}\left[\nu_{_{SC}}\right]$ (same for $\xi_{_{SC}}$), so that the effective damping will read asymptotically, e.g. for amplitude $A$ (similar for $B$),  $\left(\sigma+\text{Im}\left[\nu_{_{SC}}\right]|A|^2+\text{Im}\left[\xi_{_{SC}}\right]|B|^2\right)$.\\
\indent Hence, we do not expect the heuristic damping model to provide an accurate estimation of the actual amplitude-dependent dissipation of the system, crucial for a correct prediction of the phase-lag between forcing and the system response \citep{bauerlein2021phase}. However, accounting for a damping coefficient $\sigma$ in equations~\eqref{eq:AmpEqSCfinalA_elliptic}-\eqref{eq:AmpEqSCfinalB_elliptic} is essential in order to regularize the WNL prediction as the orbit aspect ratio $\alpha$ approaches $1$, i.g. for circular orbits, which otherwise would unrealistically predict counter-waves even for $\alpha\rightarrow 1$ \citep{faltinsen2016resonant,raynovskyy2020sloshing}, thus implying that the response curves branching is not a continuous function of $\alpha$, which is in contradiction with experimental evidences \citep{reclari2014surface}.\\

\section{Lowest order asymptotic solution}

\indent In conclusion, the lowest order asymptotic solution governing the close-to-resonance interaction of the two $m=\pm1$ counter-propagating waves is ruled by the following system of complex amplitude equations
\begin{subequations}
\begin{equation}
\label{eq:AmpEqSCfinalA_elliptic}
\frac{\text{d}A}{\text{d}t}=-\left(\sigma+\text{i}\lambda\right)A + \text{i}\,\mu \alpha_{_A} f + \text{i}\,\nu |A|^2A +\text{i}\,\xi |B|^2A,
\end{equation}
\begin{equation}
\label{eq:AmpEqSCfinalB_elliptic}
\frac{\text{d}B}{\text{d}t}=-\left(\sigma+\text{i}\lambda\right) B + \text{i}\,\mu \alpha_{_B} f + \text{i}\,\nu |B|^2B + \text{i}\,\xi |A|^2B.
\end{equation}
\end{subequations}
\noindent It follows that the leading order free surface deformation writes
\begin{eqnarray}
        \eta\left(r,\theta,t\right)=\hat{\eta}_1^{A_1,\,B_1}\left(r\right)\left(Ae^{\text{i}\left(\Omega t -\theta+\Phi_A\right)}+Be^{\text{i}\left(\Omega t +\theta +\Phi_B\right)}\right)\notag\\
        +c.c.\,.\ \ \ \ 
\end{eqnarray}
\noindent Given the mode normalization, i.e. $\hat{\eta}_1^{A_1,\,B_1}\left(r=1\right)=1$, we can express the dimensionless contact line elevation, $\delta\left(\theta,t\right)/R$, at any azimuthal coordinate, e.g. at $\theta=0$, as
\begin{equation}
    \delta\left(0,t\right)/R=\left(A+B\right)e^{\text{i}\Omega t}+c.c.\,
\end{equation}
\noindent (with $R$ the container's radius), which is the quantity that has been used for comparison with our experimental measurements.

\section{Stationary wave amplitude solutions and their stability}\label{sec:Sec2}

\begin{table*}[t!]
\centering
\begin{tabular}{c|ccccccccccc}
$H=h/R$ & 1.1 & 1.2 & 1.3 & 1.4 & 1.5 & 1.6 & 1.7 & \textbf{1.744} & 1.8 & 1.9 & 2.0 \\ \hline
$\mu$ & -0.279 & -0.280 & -0.281 & -0.282 & -0.283 & -0.283 & -0.283 & \textbf{-0.283} & -0.284 & -0.284 & -0.284 \\
$\nu$  & 1.414  & 1.407 & 1.406 & 1.407 & 1.409 & 1.410 & 1.411 & \textbf{1.412} & 1.412 & 1.412 & 1.413 \\
$\xi$   &  -7.487 & -7.914 &  -8.101 & -8.211 & -8.281 & -8.328 & -8.359 & \textbf{-8.369} & -8.381 & -8.395 & -8.405 \\ 
$\omega_0$ & 1.334 & 1.341 & 1.346 & 1.349 & 1.352 & 1.353 & 1.354 & \textbf{1.355} & 1.355 & 1.356 & 1.356 \\ \hline
\end{tabular}
\caption{Value of the normal form coefficients appearing in~\eqref{eq:AmpEqSCfinalA_elliptic}-\eqref{eq:AmpEqSCfinalB_elliptic} computed at different non-dimensional fluid depths $H=h/R$ (as reported in table~1 of \cite{marcotte2023super}) and associated with the lowest natural frequency mode. The subscript $SC$ was used in \cite{bongarzone2022amplitude} and \cite{marcotte2023super} to indicate the shape of the associated free surface response close to harmonic resonance, initially denominated single-crest (SC) by \cite{reclari2014surface}. Here the subscript SC has been omitted but in practice, $\mu$, $\nu$ and $\xi$ coincide with $\mu_{_{SC}}$, $\nu_{_{SC}}$ and $\xi_{_{SC}}$ in \cite{marcotte2023super}. For completeness we also report the value of the system's lowest natural frequency $\omega_0$. The bold values correspond to those used in the main document for comparison with experiments.}
\label{tab:Tab1}
\end{table*}

\noindent By turning~\eqref{eq:AmpEqSCfinalA_elliptic}-\eqref{eq:AmpEqSCfinalB_elliptic} into polar coordinates, i.e. $A=|A|e^{\text{i}\Phi_A}$ and $B=|B|e^{\text{i}\Phi_B}$, we can split real and imaginary parts, hence obtaining
\begin{subequations}
\begin{equation}
\label{eq:SysEqHarm_Gen_A}
\frac{\text{d}|A|}{\text{d}t}=-\sigma |A| + \alpha_{_A}\mu f\sin{\Phi_A},
\end{equation}
\begin{equation}
\label{eq:SysEqHarm_Gen_PhiA}
|A|\frac{\text{d}\Phi_A}{\text{d}t}=-\lambda |A|+\alpha_{_A}\mu f\cos{\Phi_A}+\nu |A|^3+\xi |B|^2|A|,
\end{equation}
\begin{equation}
\label{eq:SysEqHarm_Gen_B}
\frac{\text{d}|B|}{\text{d}t}=-\sigma |B| + \alpha_{_B}\mu f\sin{\Phi_B},
\end{equation}
\begin{equation}
\label{eq:SysEqHarm_Gen_PhiB}
|B|\frac{\text{d}\Phi_B}{\text{d}t}=-\lambda |B|+\alpha_{_B}\mu f\cos{\Phi_B}+\nu |B|^3+\xi |A|^2|B|.
\end{equation}
\end{subequations}
\noindent Let us then decompose amplitudes and phases as the sum of stationary values plus time-dependent small perturbations of order $\epsilon\ll1$.
\begin{eqnarray}
\label{eq:LinHarm_AB_PhiAB}
\mathbf{y}\left(t\right)=\left(
\begin{matrix}
|A|\left(t\right) \\
\Phi_A\left(t\right)\\
|B|\left(t\right)\\
\Phi_B\left(t\right)
\end{matrix}
\right)=\left(
\begin{matrix}
a_0\\
\phi_{A,0}\\
b_0\\
\phi_{B,0}
\end{matrix}
\right)+\epsilon\left(
\begin{matrix}
a_1\left(t\right)\\
\phi_{A,1}\left(t\right)\\
b_1\left(t\right)\\
\phi_{B,1}\left(t\right)
\end{matrix}
\right)=\notag \\
=\mathbf{y}_0+\epsilon\mathbf{y}_1\left(t\right)=\mathbf{y}_0+\epsilon\left(\hat{\mathbf{y}}_1e^{s t}+\text{c.c.}\right),\ \ \ 
\end{eqnarray}
\noindent with $s=s_{_R}+\text{i}s_{_I}\in\mathbb{C}$ an eigenvalue and $\text{c.c.}$ denoting the complex conjugate part of the small linear perturbation. The ubstitution of~\eqref{eq:LinHarm_AB_PhiAB} in~\eqref{eq:SysEqHarm_Gen_A}-\eqref{eq:SysEqHarm_Gen_PhiB} and the linearization around $\mathbf{y}_0$, lead to two problems at order $\epsilon^0$ and $\epsilon$, respectively. As the nonlinear system of equations at order $\epsilon^0$ does not admit an analytical solution, we apply a numerical procedure after rewriting the problem in the form:
\begin{equation}
\label{eq:eps0mat_harm}
\mathcal{F}=0=\begin{cases}
\alpha_{_A}\mu f\sin{\phi_{A,0}} -\sigma a_0,\\
\alpha_{_A}\mu f\cos{\phi_{A,0}} - a_0\left(\lambda -\nu a_0^2-\xi b_0^2\right),\\
\alpha_{_B}\mu f\sin{\phi_{B,0}} -\sigma b_0,\\
\alpha_{_B}\mu f\cos{\phi_{B,0}} - b_0\left(\lambda-\nu b_0^2-\xi a_0^2\right).
\end{cases}
\end{equation}
\noindent System~\eqref{eq:eps0mat_harm} is then solved in Matlab function using the built-in function \textit{fsolve} and prescribing some initial guesses (\textit{ig}) for $\left(a_0^{ig},\phi_{A,0}^{ig},b_0^{ig},\phi_{B,0}^{ig}\right)$. In practice, we provide in input the external control parameters, $\Omega$, $a_x=\overline{a}_x/R$ and $\alpha$, whereas the associated combination of stationary amplitudes and phases, $\left(a_0,\phi_{A,0},b_0,\phi_{B,0}\right)$ are computed as outputs.\\
\indent In the following we study the stability properties of these steady-state amplitude and phase solutions. Given the ansatz $\mathbf{y}_1\left(t\right)=\hat{\mathbf{y}}_1e^{st}+c.c.$, at order $\epsilon$ the linearized and unsteady system, describing the evolution of small amplitude perturbations around the stationary states, can be written in a matrix form as
\begin{equation}
\label{eps1mat_harm_mat}
s\mathbf{M}\hat{\mathbf{y}}_1=\mathbf{K}\hat{\mathbf{y}}_1,
\end{equation}\\
\noindent with matrices $\mathbf{M}$ and $\mathbf{K}$ reading
\begin{equation}
\label{eq:eps1_mat_MK}
\mathbf{M}=\left[
\begin{matrix}
1 & 0 & 0 & 0\\
0 & a_0 & 0 & 0 \\
0 & 0 & 1 & 0\\
0 & 0 & 0 & b_0
\end{matrix}
\right],\ \ \ \mathbf{K}=\left[
\begin{matrix}
K_{11} & K_{12} & 0 & 0\\
K_{21} & K_{22} & K_{24} & 0 \\
0 & 0 & K_{33} & K_{34}\\
K_{41} & 0 & K_{43} & K_{44}
\end{matrix}
\right],
\end{equation}
\noindent $\hat{\mathbf{y}}_1=\left(\hat{a}_1,\hat{\phi}_{A,1},\hat{b}_1,\hat{\phi}_{B,1}\right)^T$ and 
\begin{subequations}
\begin{equation}
\label{eq:eps1_mat_K_11_33}
K_{11} =-\sigma,\ K_{33} =-\sigma,
\end{equation}
\begin{equation}
\label{eq:eps1_mat_K_12_34}
K_{12} =\alpha_{_A}\mu f\cos{\phi_{A,0}},\ K_{34} =\alpha_{_B}\mu f\cos{\phi_{B,0}},
\end{equation}
\begin{equation}
\label{eq:eps1_mat_K_21_43}
K_{21} =-\lambda+3\nu a_0^2+\xi b_0^2,\ K_{43} =-\lambda+3\nu b_0^2+\xi a_0^2,
\end{equation}
\begin{equation}
\label{eq:eps1_mat_K_22_44}
K_{22} =-\alpha_{_A}\mu f\sin{\phi_{A,0}},\ K_{44} =-\alpha_{_B}\mu f\sin{\phi_{B,0}},
\end{equation}
\begin{equation}
\label{eq:eps1_mat_K_24_41}
K_{24} =2\xi a_0 b_0,\ K_{41} =2\xi a_0 b_0,
\end{equation}
\end{subequations}
\noindent We proceed as follows. For each $\left(a_0,\phi_{A,0},b_0,\phi_{B,0}\right)$, solution of~\eqref{eq:eps0mat_harm}, we obtain four eigenvalues $s$. If the real part of at least one of these eigenvalues is positive, then that configuration, associated with the set of external parameters $\left(\Omega,a_x,\alpha\right)$, is labeled as unstable.

\section{Values of the normal form coefficients}\label{sec:Sec4}

In table~\ref{tab:Tab1} we report the values of the normal form coefficients, $\mu$, $\nu$ and $\xi$ appearing in~\eqref{eq:AmpEqSCfinalA_elliptic}-\eqref{eq:AmpEqSCfinalB_elliptic} as a function of the non-dimensional fluid depth $H=h/R$. Experiments reported in the main document have been performed at a fluid depth $H=1.744$.\\
\indent For completeness we also report the value of the system's lowest natural frequency $\omega_0$, which satisfies the well-known dispersion relation for gravity waves $\omega_0=\overline{\omega}_0/\sqrt{g/R}=\sqrt{k\tanh{\left(kH\right)}}$ (with $k=1.8412$) \cite{Lamb32}. We do not report the value of the damping coefficient $\sigma$ as a function of $H$, since for $H\ge1$ the fluid depth does not affect significantly its value estimated according to~\eqref{eq:dampM}.

\bibliography{Bibliography}